\documentstyle[psfig]{aa}

\def\hour{\hbox{$^{h}$}}

\def\min{\hbox{$^{m}$}}

\def\degr{\hbox{$^{\mbox{\scriptsize o}}$}}

\def\arcmin{\hbox{$^\prime$}}

\begin{document}

\thesaurus{}

\title{Search for young stars among ROSAT All-Sky Survey 
X-ray sources in and around the R CrA dark cloud \thanks{Partly based on 
observations collected at the 1.52m and 3.5m telescopes of the European 
Southern Observatory, Chile, in programs 55.E-0549, 57.E-0646, and 63.L-0023,
and on observations collected at the 0.9m, 1.5m, and 4.0m CTIO telescope.} }

\author{R. Neuh\"auser\inst{1}, F.M. Walter\inst{2}, E. Covino\inst{3}, 
J.M. Alcal\'a\inst{3}, S.J. Wolk\inst{4}, S. Frink\inst{5}, 
P. Guillout\inst{6}, M.F. Sterzik\inst{7}, F. Comer\'on\inst{8} }

\offprints{Ralph Neuh\"auser}

\institute{ Max-Planck-Institut f\"ur extraterrestrische Physik, D-85740 Garching,
Germany, rne@mpe.mpg.de 
\and
Department of Physics and Astronomy, SUNY, Stony Brook, NY 11794-3800, USA
\and
Osservatorio Astronomico di Capodimonte, I-80131 Napoli, Italy
\and
Harvard-Smithonian Center for Astrophysics, Cambridge MA 02138, USA
\and
University of California San Diego, La Jolla, CA 92093, USA
\and
Observatoire Astronomique, CNRS UMR 7550, F-67000 Strasbourg, France
\and
European Southern Observatory, Casilla 19001, Santiago 19, Chile
\and
European Southern Observatory, Karl-Schwarzschild-Stra\ss e 2, D-85748 Garching, Germany
}

\date {Received March 2000; accepted 2 August 2000}

\maketitle

\markboth{Neuh\"auser et al.: Young stars in and around 
R CrA}{Neuh\"auser et al.: Young stars in and around R CrA}

\begin{abstract}
We present the ROSAT All-Sky Survey data in 
a 126 deg$^{2}$ area in and around the CrA star forming region.
With low-resolution spectroscopy of unidentified ROSAT sources
we could find 19 new pre-main sequence stars,
two of which are classical T~Tauri stars, the others being weak-lined.
The spectral types of these new T~Tauri stars range from F7 to M6.
The two new classical T~Tauri stars are located towards 
two small cloud-lets outside of the main CrA cloud.
They appear to be $\sim 10$ Myrs old, by comparing their location in 
the H-R diagram with isochrones for an assumed distance of 130 pc,
the distance of the main CrA dark cloud.
The new off-cloud weak-line T~Tauri stars may have formed in 
similar cloud-lets, which have dispersed recently.
High-resolution spectra of our new T~Tauri stars show that they have
significantly more lithium absorption than zero-age main-sequence stars 
of the same spectral type, so that they are indeed young.
From those spectra we also obtained rotational and radial velocities.
For some stars we found the proper motion in published catalogs.
The direction and velocity of the 3D space motion -- south relative to the 
galatic plane -- of the CrA T~Tauri stars is consistent with the 
dark cloud being formed originally by a high-velocity cloud impact onto 
the galactic plane, which triggered the star formation in CrA.
We also present VRIJHK photometry for most of the new 
T~Tauri stars to derive their luminosities, ages, and masses.


\end{abstract}

\section {Introduction}
\label{intro}

The Corona Australis (CrA) molecular cloud complex (Dame et al. 1987) is one of the 
nearest regions of ongoing and/or recent intermediate- and low-mass star formation. 
The dark cloud near the emission line star R CrA (Knacke et al. 1973)
is the densest cloud core with extinction up to $A_{V} \sim 45$ mag 
(Wilking et al. 1992). This cloud is also called FS~445-47 
(Feitzinger \& St\"uwe 1984) and {\em condensation A} (Rossano 1978);
Harju et al. (1993) resolved cloud {\em A} into five condensations,
the star R~CrA being located in {\em A2}.
Cambr\'esy (1999) mapped the cloud using optical star counts.
Between the stars R and T~CrA, there is the reflection nebula NGC~6729;
TY~CrA and HD 176386 illuminate the nebula NGC~6726/6727.
Several infrared (IR) surveys revealed a large population of
embedded IR sources (Taylor \& Storey 1984, Wilking et al. 1984, 1986, 1992, 1997),
some of which are IR Class I objects, extremely young stars still deeply embedded 
in their dense circumstellar envelopes (Adams et al. 1987, 
Andr\'e \& Montmerle 1994).
From the main-sequence contraction time of the early-type stars 
R, T, and TY~CrA, the age of the cloud is estimated to be between 
$\le 1$ (Knacke et al. 1973) and $6$~Myrs (Wilking et al. 1992).

The distance towards the CrA star forming region was estimated
by Gaposchkin \& Greenstein (1936) to be $150 \pm 50$~pc and later 
by Marraco \& Rydgren (1981) to be $\sim 129$~pc (assuming R=4.5).
The Hipparcos satellite tried to measure the parallax of the 
star R CrA and found $122 \pm 68$ mas, i.e. no reliable solution.
Casey et al. (1998) estimated the distance towards the 
eclipsing double-lined spectroscopic binary TY CrA
to be $129 \pm 11$~pc from their orbit solution.

Only a few low-mass pre-main sequence (PMS) stars, so-called T~Tauri stars (TTS), 
associated with the CrA dark cloud had been found by H$\alpha$ and IR surveys 
(Knacke et al. 1973, Glass \& Penston 1975, Marraco \& Rydgren 1981, 
Wilking et al. 1984, 1986, 1992, 1997), all being classical TTS (cTTS)
with IR excess and strong H$\alpha$ emission (see Table 1).
Patten (1998) obtained optical photometry and spectroscopy of some more
sources of, by then, unknown nature around R CrA, previously found by
Knacke et al. (1973), Glass \& Penston (1975), and Marraco \& Rydgren (1981),
and of some X-ray sources found in a pointed ROSAT observation. He
classified some of them as new association members 
due to H$\alpha$ emission.

Table 1 gives a list of all previously known optically visible young stars in CrA,
with their names, PMS types (Herbig Ae/Be or T Tauri stars), spectral types,
H$\alpha$ and lithium equivalent widths, and some remarks, e.g. on radial
velocity and binarity.

\begin{table*}
\label{preTTS}
\caption []{ {\bf Previously known or suspected optically visible young stars in CrA.} \\
PMS types are either cTTS or wTTS or intermediate-mass Herbig Ae/Be stars, 
one being of spectral type F0e. We also list spectral types, H$\alpha$ and 
lithium 6708\AA~equivalent widths (negative when in emission) 
as well as radial velocities (RV in $km~s^{-1}$), if available.
Data for stars with CrAPMS designations are from Walter et al. (1997), 
for other stars with HBC number from the Herbig-Bell catalog 
(HBC, Herbig \& Bell 1988), unless otherwise noted. 
At the end of the table, we also list four more late B-type stars, 
which might be associated with the CrA cloud.
Some of the previously suspected TTS have been confirmed by our spectroscopy,
but for MR81 H$\alpha$ 10 and Kn anon 2, we could not detect lithium.
}

\vspace{-0.5cm}

\begin{tabular}{rlllllrll} \\ \hline
No. & Designation & GP75 & Other name & PMS & Spec & $W_{\lambda}(H\alpha )$ 
& $W_{\lambda}$(Li) & Remarks \\
{\scriptsize HBC} & & name & & type & type & [\AA ] & [\AA ] & \\ \hline

286 & S CrA    & & Hen 3-1731      & cTTS & K6    & $-90.0$ & 0.39$^{e}$ & RV=0, 1.37" binary \\ 
287 & TY CrA   & & CrAPMS 11       & HeBe & B9e$^{a}$&      & yes & triple, d=129~pc$^{b}$ \\
288 & R CrA    & & CoD$-37^{\circ}13027$ & HeAe & A5e II &  & & RV=$-36.0$ \\
289 & DG CrA   & & Hen 3-1734      & cTTS & K0$^{e}$  & $-77.9^{e}$ & 0.57$^{e}$ & \\
290 & T CrA    & &                 & HeFe & F0e             & & & \\
291 & VV CrA   & & Hen 3-1736      & cTTS & K1$^{e}$     & $-72.0^{e}$ & & \\
673 & MR81 H$\alpha$ 10 & &          &      & K            & $1.0 ^{e}$  & no$^{e}$ & non-TTS$^{e,g}$ \\
674 & MR81 H$\alpha$ 6  & & CrAPMS 7 & cTTS & M1              & $-33.5$ & 0.36 & \\ 
675 & Kn anon 2  & j2 & &      & G0$^{e}$  & $1.0^{e}$ & no$^{e}$ & non-TTS$^{e,g}$ \\
676 & CoD$-37^{\circ}13022$ & i2 & CrAPMS 1 & wTTS & K1      & $  0.3$ & 0.39 & RV=$-1.0$ \\
677 & MR81 H$\alpha$ 2 & i & HaGr 1-100    & cTTS$^{e}$ & K8        & $-46.0^{e}$ & $0.47^{e}$ & \\
678 & V702 CrA & a2 & CrAPMS 2        & wTTS & G5            & $ -0.7$ & 0.28 & RV=$-1.2$ \\
679 & CrAPMS 3 & w &            & wTTS & K2    & $-0.9$     & 0.41 & RV=$-1.2$, 4.5" pair \\
    & CrAPMS 3/c & & & wTTS$^{e}$ & M4$^{e}$ & $-6.8^{e}$ & $0.36^{e}$ & 4.5" pair \\
680 & MR81 H$\alpha$ 14 & &        & wTTS$^{e}$ & M3$^{e}$ & $-4.6^{e}$ & 0.64$^{e}$ & TTS$^{g}$ \\
    & & e2  &       &            & M3-5$^{g}$ & em.$^{g}$ & & TTS$^{g}$ \\
    & & g2  &       &            & M3-5$^{g}$ & em.$^{g}$ & & TTS$^{g}$ \\
    & & f2  &       &            & K4$^{g}$ &             & & TTS$^{g}$ \\
    & & n   &       & cTTS$^{g}$ &          & em.$^{g}$ & & \\
    & MR81 H$\alpha$ 12 & &          &          & M3-5$^{g}$ & em.$^{g}$ & & TTS$^{g}$ \\
    & MR81 H$\alpha$ 13 & &          &          & M3-5$^{g}$ & em.$^{g}$ & & TTS$^{g}$ \\
    & MR81 H$\alpha$ 15 & &          &          & M3-5$^{g}$ & em.$^{g}$ & & TTS$^{g}$ \\
    & MR81 H$\alpha$ 16 & &          &          & M1$^{g}$ & em.$^{g}$ & & TTS$^{g}$ \\
    & MR81 H$\alpha$ 17 & &          &          & M3-5$^{g}$ & em.$^{g}$ & & TTS$^{g}$ \\
    & CrAPMS 4NW & & & wTTS & M0.5  & $-1.1$  & 0.45 & RV=$-2.2$ \\
    & CrAPMS 4SE & & & wTTS & G5  & $1.0$   & 0.36 & RV=$-2.0$ \\
    & CrAPMS 5   & & & wTTS & K5  & $-0.8$  & 0.44 & RV=$-0.8$ \\
    & MR81 H$\alpha$ 11NE & & CrAPMS 6NE & wTTS & M3    & $-5.9$  & 0.70 & NE/SW 3" pair \\
    & MR81 H$\alpha$ 11SW & & CrAPMS 6SW & wTTS & M3.5  & $-9.8$  & 0.44 & NE/SW 3" pair\\
    & CrAPMS 8   & g2 & & wTTS & M3    & $-3.9$  & 0.57 & \\
    & CrAPMS 9   & & & wTTS & M2    & $-9.2$  & 0.5  & \\ 
& RXJ1855.1-3754 & & GSC 07916-00050 & wTTS$^{d}$ & K3$^{d}$ & $1.6^{d}$ & $0.38^{d}$ & $W$(Li)$\simeq W$(Ca)$^{d}$ \\
& RXJ1857.7-3719 & & Patten R1c  & & M3-5$^{g}$ & em.$^{g}$ & & TTS$^{g}$ \\
& RXJ1858.9-3640 & & Patten R17c & & M3-5$^{g}$ & em.$^{g}$ & & TTS$^{g}$ \\
& RXJ1859.7-3655 & & Patten R13a & & M3$^{g}$   & em.$^{g}$ & & TTS$^{g}$ \\ \hline

 & HR 7169    & l & HD 176269 & & B9 & 7.7$^{e}$ & & $d \simeq 134$~pc$^{f}$ (*) \\ 
 & HR 7170    & k & HD 176270 & & B8 & 7.0$^{e}$ & & $d \simeq 77$~pc$^{f}$ (*) \\
 & SAO 210888 & & HD 177076 & & B9.5  &  & & $d \simeq 185$~pc$^{f}$ \\ 
 & HD 176386  & p & HIP 93425 & & B9 & & & \hspace{-2cm} RV=7.3$^{c}$, $d \simeq 136$~pc$^{f}$, 4" binary$^{c}$ \\ \hline

\end{tabular}

References: (a) HBC, (b) Casey et al. 1998, (c) Simbad, (d) Neuh\"auser et al. 
1997, (e) this work, (f) Hipparcos, (g) Patten 1998. \\
Note:
%
%
(*) Both HR 7169 and HR 7170 are spectroscopic binaries (Hoffleit 1982).
With a separation of 13" in 1982.66 (Torres 1985), this visual pair may be bound.
The system was also detected by EO as the spatially unresolved
source PMSCrA 10 (Walter et al. 1997).

\end{table*}

With optical follow-up observations of previously unidentified X-ray sources
detected with the {\em Einstein Observatory} (EO), Walter (1986) and 
Walter et al. (1997) found eleven new TTS members, namely CrAPMS 1 to 9,
two of them (CrAPMS 4 and 6) being visual pairs consisting of two PMS stars
(see Table 1). With only one exception (the cTTS CrAPMS 7), all 
of them are weak-emission line TTS\footnote{
Because most wTTS lack the IR excess typical for cTTS and, hence, are 
probably not surrounded by circumstellar disks, Walter (1986) called
them naked TTS. However, because wTTS and nTTS populate both the 
convective and radiative PMS tracks, the wTTS and nTTS population
is different from the so-called post-TTS population. Post-TTS are 
older than cTTS and do not show signatures of circumstellar disks 
and accretion (H$\alpha$, UV, IR); post-TTS are expected to exist in large 
numbers, if star formation has been going on for longer than the typical 
life-time of circumstellar disks (Herbig 1978).} (wTTS).

In Table 1 we list all the previously known and suspected young stars in CrA,
which are optically visible.
Walter et al. (1997) could also establish the typical radial velocity
of kinematic members of the CrA association: All the seven TTS, for
which radial velocities are known, show velocities in the range of 
$-2$ to $0~km~s^{-1}$ (heliocentric).
Chen et al. (1997) also compiled a list of young stars in CrA and
estimated their bolometric luminosities. 
We omit IR Class I sources and brown dwarf candidates in this paper,
because they are too faint in X-rays to be detected 
in the ROSAT All-Sky Survey (RASS).

The early-type stars TY CrA, HR 7169, and HR 7170, all being spectroscopic
binaries, were also detected by EO (Walter et al. 1997), but their X-ray 
emission may originate from late-type companions. 
While many of the optically visible TTS are known to
be rather strong and variable X-ray emitters (eg. Montmerle et al. 1983,
Walter et al. 1988, Neuh\"auser et al. 1995), it was surprising that
a few IR Class I objects were also detected by ASCA and ROSAT X-ray
observations (Koyama et al. 1996, Neuh\"auser \& Preibisch 1997).
Wilking et al. (1997) also found five brown dwarf candidates, but they
are not detected in deep ROSAT pointings (Neuh\"auser et al. 1999).

Because there are several early-type stars in the CrA association, 
there should be much more than the $\sim 3$ dozen TTS listed in
Table 1, if its initial mass function (IMF) is consistent 
with the Miller-Scalo IMF (Miller \& Scalo 1979).
From the spatial incompleteness of the EO observations and the 
X-ray variability of TTS, Walter et al. (1997) concluded that there
should be $\sim 70$ TTS in CrA.
From their IR survey, Wilking et al. (1997) estimated the number of the
low-mass members 
to be 22 to 40 for an association age of $\sim 3$ Myrs.
If star formation has been ongoing in CrA for more than $\sim 3$ Myrs,
there should be even more PMS stars. Such
older PMS stars, i.e. the post-TTS, should partly be found around
the CrA dark cloud, because they had enough time to disperse out.

Optical follow-up observations of RASS sources
in and around other star forming regions (Tau-Aur, Orion, Cha, $\rho$ Oph, 
Lup-Sco-Cen, etc.) revealed large populations of previously unknown 
PMS stars most of them being wTTS (see Neuh\"auser 1997 for a review),
identified as such with low- to intermediate resolution spectroscopy
showing late spectral types, H$\alpha$ emission (or emission filling-in
the absorption), and lithium 6708\AA~absorption, a youth indicator.
Because some of them were found even outside the star forming clouds,
it was argued (Brice\~no et al. 1997) that many of these young stars 
are not PMS, but zero-age main-sequence (ZAMS) stars similar 
to the Pleiades, which also show strong X-ray emission, H$\alpha$ 
absorption, and lithium 6708\AA~absorption. However, in the meantime,
Covino et al. (1997), Neuh\"auser et al. (1997), Wichmann et al. 
(1999), and Alcal\'a et al. (2000) have shown with high-resolution
spectra that most of the previously claimed wTTS really are PMS stars,
because they show more lithium than ZAMS stars of the same spectral type.
Also, Neuh\"auser \& Brandner (1997) found that all 15 Li-rich stars
found by ROSAT, which could be placed accurately into the H-R diagram 
using Hipparcos data, clearly are PMS stars.
The young stars newly found outside of the clouds could either 
be ejected out of their parent cloud (Sterzik \& Durisen 1995), or
they could have formed locally in small cloud-lets which dispersed 
since then (Feigelson 1996, Mizuno et al. 1998).
Many of the new ROSAT TTS in Taurus, Orion, and Lup-Sco-Cen are probably
members of the Gould Belt (Guillout et al. 1998a,b), young stars still
at least slightly above the ZAMS.

The CrA dark cloud is located $\sim 18^{\circ}$ below the galactic plane.
According to Olano (1982, see also Fig. 1.10 in P\"oppel 1997), the CrA 
and Chamaeleon clouds are not part of the Gould Belt nor
of the Lindblad ring, because both CrA and Cha are far below the galactic
plane, while the belt and the ring are both above the plane. 
In the cross-correlation of Tycho and RASS, the CrA association is seen 
as a small cluster of X-ray active stars around $l\simeq 0^{\circ}$ and 
$b \simeq -20 ^{\circ}$, see Fig. 3 in Guillout et al. (1998a) and 
Fig. 3 in Guillout et al. (1998b). The Gould Belt is above the galactic plane
in this quadrant. Hence, neither CrA nor Cha are part of the belt,
so that we should not expect to detect Gould Belt members in CrA.
Many new TTS were discovered around the Cha clouds (Alcal\'a et al. 1995,
Covino et al. 1997), so that we may expect to find many such TTS also
here around the CrA dark cloud.
Hence, we carried out an optical identification program to find more PMS 
in and around the CrA dark cloud among unidentified 
RASS sources.\footnote
{Four new PMS stars were already identified among ROSAT sources:
GSC 07916-00050 as the optical counterpart to RXJ1855.1-3754 (Neuh\"auser et al. 1997),
an X-ray source found with the ROSAT High Resolution Imager (HRI) in a deep pointed 
observation centered on RXJ1856.5-3754, an isolated radio-quiet neutron 
star (Walter et al. 1996, Walter \& Matthews 1997). Neuh\"auser et al. (1997)
identified GSC 07916-00050 in order to confirm that it is the true 
(i.e. possibly X-ray emitting) counterpart to RXJ1855.1-3754, to be able to 
perform a correct boresight correction of the position of the central HRI source.
In addition, Patten (1998) identified three new PMS stars among X-ray sources 
found in a pointed observation with the ROSAT Positional Sensitive Proportional
Counter (PSPC), also listed in Table 1.}
In Sect. \ref{xrays}, we describe the X-ray data reduction and list all X-ray sources
found with RASS (Table 2). Our spectra are presented in Sect. \ref{ocp} together with 
lists of potential optical counterparts (Table 3). 
Then, in Sect. \ref{resspec}, we discuss the results of the
spectroscopy including Table 4 with our new TTS.
In Sect. \ref{yso} we list the available optical and IR 
photometry for the new TTS;
the H-R diagram is shown and discussed in Sect. 6.
Then, in Sect. \ref{pmr}, we present proper motions of some of our
new PMS stars and discuss the 3D space motion of young stars in CrA.
Finally, in Sect. \ref{comp}, we estimate the completeness of our survey.
We summarize our results in the last section.

\section {X-ray data reduction }
\label{xrays}

The previously known young stars associated with the CrA dark cloud
are located in a small, $\sim 126~deg ^{2}$ area around R~CrA 
near $\alpha _{2000} = 19^{h}$ and $\delta _{2000} = -37 ^{\circ}$.
Because we also expect to find previously unknown young stars around
the dark cloud, we investigated the area
$\alpha _{2000} = 18^{h}~35^{m}$ to $19^{h}~38^{m}$  
and $\delta _{2000} = -41^{\circ}$ to $-33^{\circ}$.

We reduced all RASS II data\footnote
{Prior to the optical follow-up observations, we reduced the earlier version of
the ROSAT data (RASS I) with an earlier version of EXSAS; although the source 
lists and X-ray properties are sightly different, all newly identified young stars are
detected in both RASS I and RASS II.} pertaining to that area with
the Extended Scientific Analysis Software (EXSAS, Zimmermann et al. 1994)
version 98APR running under ESO-MIDAS version 97NOV.
We performed standard local and map source detection in the bands
soft ($0.1$ to $0.4~keV$), hard 1 ($0.5$ to $0.9~keV$), hard 2 ($0.9$ to $2.0~keV$),
hard ($0.5$ to $2.0~keV$), and broad ($0.1$ to $2.4~keV$), as described in detail
in Neuh\"auser et al. (1995).
After merging the source lists, each source was again tested in the above 
mentioned five bands by a maximum likelihood source detection algorithm.
Following Neuh\"auser et al. (1995) we accept only sources with 
$ML \ge 7.4$ corresponding to $\sim 3.5~\sigma$ as real.

The moderate energy resolution of the PSPC allows to extract some spectral
information from the RASS data, namely so-called hardness ratios, 
i.e., X-ray colors, which are defined as follows:
If $Z_{s,h1,h2}$ are the count rates in the bands soft, hard 1, 
and hard 2, respectively, then 
\begin{equation}
\label{hr}
HR~1~=~\frac{ Z_{h1} + Z_{h2} - Z_{s} } { Z_{h1} + Z_{h2} + Z_{s} }
\quad \mbox{\&} \quad
HR~2~=~\frac{ Z_{h2} - Z_{h1} } { Z_{h2} + Z_{h1} }
\end{equation}
I.e., hardness ratios range from $-1$ to $+1$, 
hardness ratio errors can be larger than $2$.
If no counts are detected, e.g, in the soft band, 
then HR~$1~=~1$, but one can estimate a lower 
limit to HR~$1$ by using the upper limit to the soft 
band count rate $Z_{s}$ in Eq. (\ref{hr}),
similar for upper limits to HR~$1$ as well as for 
upper and lower limits to HR~$2$.

\begin{table*}
\label{tab2x}
\caption []{ {\bf X-ray data for our sample.} \\
Listed are all X-ray sources
detected in the ROSAT All-Sky Survey sorted by right ascension.
We list a running number (used in other tables for cross-reference),
the ROSAT source designation, the X-ray source position ($J2000.0$),
the maximum likelihood of existence $ML$, the hardness ratios with errors,
the (background subtracted and vignetting corrected) number of counts in the 
broad band ($0.1$ to $2.4~keV$) with errors, the exposure time in seconds, 
and $\log ECF$ for the energy conversion factor in counts $erg^{-1}~cm^{2}$.}

\begin{tabular}{rlccrrrr@{$\pm$}lrr}
\hline
No. & Designation & \multicolumn{2}{c}{$\alpha _{2000}$ (X-ray) $\delta _{2000}$}
 & $ML$ & HR~1 & HR~2 & \multicolumn{2}{c}{Counts} & \hspace{-.2cm} Exp. [s] & \hspace{-.2cm} $\log ECF$\\ \hline

  1&RXJ1835.1-3404 & 18:35:08.0 & -34:04:20.0 &  10.2&$     \ge  0.61$&$ 0.30 \pm 0.36$&  8.9& 4.0&147.2&-10.87 \\
  2&RXJ1835.3-3927 & 18:35:21.5 & -39:27:13.3 &   8.1&$ 0.15 \pm 0.49$&$     \ge  0.10$&  7.1& 3.2&124.3&-11.04 \\
  3&RXJ1835.7-3259 & 18:35:47.0 & -32:59:41.5 &1240.1&$ 0.93 \pm 0.02$&$ 0.53 \pm 0.05$&364.0&19.6&163.0&-10.88 \\
  4&RXJ1835.8-3813 & 18:35:49.2 & -38:13:24.1 &   9.8&$ 0.77 \pm 0.41$&$ 0.02 \pm 0.44$& 11.7& 4.3&139.4&-10.91 \\
  5&RXJ1835.8-4046 & 18:35:53.4 & -40:46:23.5 &  10.2&$     \ge  0.27$&$     \ge  0.43$&  3.6& 2.3&125.7&-10.87 \\
  6&RXJ1835.9-3525 & 18:35:58.3 & -35:25:35.5 &   7.7&$ 0.77 \pm 0.44$&$     \ge  0.42$&  6.1& 3.4&143.3&-10.91 \\
  7&RXJ1836.3-4010 & 18:36:18.7 & -40:10:20.2 &  13.1&$-0.65 \pm 0.40$&$     \le  0.19$& 11.2& 4.0&131.5&-11.31 \\
  8&RXJ1836.6-3451 & 18:36:40.6 & -34:51:33.5 &  13.2&$     \ge  0.44$&$ 0.28 \pm 0.42$&  9.4& 3.9&139.1&-10.87 \\
  9&RXJ1837.3-3442 & 18:37:18.2 & -34:42:39.0 &  13.1&$ 0.25 \pm 0.40$&$ 0.05 \pm 0.54$&  9.2& 3.5&133.2&-11.02 \\
 10&RXJ1838.2-3838 & 18:38:17.1 & -38:38:12.9 &  11.8&$     \ge  0.47$&$ 0.31 \pm 0.41$&  8.0& 3.3&142.1&-10.87 \\
 11&RXJ1838.3-3523 & 18:38:20.2 & -35:23:32.6 &  10.3&$-0.67 \pm 0.52$&$     \le  0.42$&  9.4& 3.8&140.0&-11.32 \\
 12&RXJ1839.0-3705 & 18:39:05.8 & -37:05:46.8 &  21.6&$ 0.84 \pm 0.24$&$-0.03 \pm 0.31$& 13.4& 4.2&139.0&-10.89 \\
 13&RXJ1839.0-3726 & 18:39:05.9 & -37:26:36.0 &  12.3&$ 0.78 \pm 0.21$&$-0.30 \pm 0.39$&  9.7& 3.9&142.1&-10.90 \\
 14&RXJ1839.2-3854 & 18:39:16.1 & -38:54:13.8 &   9.2&$     \ge  0.18$&$     \ge  0.40$&  3.3& 2.5&135.4&-10.87 \\
 15&RXJ1839.7-3458 & 18:39:46.8 & -34:58:58.2 &   7.8&$     \ge  0.55$&$ 0.51 \pm 0.35$&  5.8& 3.2&167.4&-10.87 \\
 16&RXJ1840.1-3640 & 18:40:09.5 & -36:40:09.1 &   7.9&$ 0.24 \pm 0.40$&$ 0.08 \pm 0.52$&  8.6& 3.6&154.6&-11.02 \\
 17&RXJ1840.6-3728 & 18:40:38.3 & -37:28:44.3 &  13.0&$     \ge  0.67$&$ 0.29 \pm 0.36$&  8.6& 3.6&150.6&-10.87 \\
 18&RXJ1840.6-3612 & 18:40:39.7 & -36:12:14.1 &   7.8&$ 0.34 \pm 0.37$&$-0.01 \pm 0.48$& 11.1& 4.2&150.4&-10.99 \\
 19&RXJ1840.8-3547 & 18:40:53.8 & -35:47:06.2 &  32.7&$-0.13 \pm 0.25$&$-0.19 \pm 0.38$& 20.2& 5.0&147.3&-11.12 \\
 20&RXJ1840.9-3350 & 18:40:55.1 & -33:50:52.5 &  14.9&$ 0.28 \pm 0.30$&$ 0.38 \pm 0.33$& 16.4& 5.1&180.2&-11.01 \\
 21&RXJ1841.2-3821 & 18:41:13.5 & -38:21:55.9 &   9.1&$     \le -0.49$&$              $&  5.7& 3.1&151.9&-11.52 \\
 22&RXJ1841.5-3508 & 18:41:34.2 & -35:08:27.7 &  11.2&$ 0.15 \pm 0.38$&$-0.28 \pm 0.53$&  9.5& 3.8&154.7&-11.04 \\
 23&RXJ1841.8-3525 & 18:41:49.7 & -35:25:44.3 &  58.2&$ 0.47 \pm 0.18$&$-0.14 \pm 0.24$& 26.6& 5.6&140.0&-10.97 \\
 24&RXJ1842.1-3732 & 18:42:12.0 & -37:32:33.0 &   9.4&$ 0.05 \pm 0.38$&$-0.09 \pm 0.53$&  9.7& 3.8&141.1&-11.07 \\
 25&RXJ1842.8-3845 & 18:42:48.6 & -38:45:28.3 &   8.7&$     \ge  0.40$&$ 0.46 \pm 0.45$&  8.7& 3.6&137.4&-10.87 \\
 26&RXJ1842.9-3451 & 18:42:58.2 & -34:51:00.3 &   7.6&$ 0.95 \pm 0.50$&$     \ge  0.22$&  3.7& 2.4&154.3&-10.87 \\
 27&RXJ1842.9-3532 & 18:42:58.4 & -35:32:34.6 &  21.7&$ 0.90 \pm 0.24$&$ 0.29 \pm 0.34$& 10.9& 3.9&135.3&-10.88 \\
 28&RXJ1843.0-3331 & 18:43:05.2 & -33:31:07.6 &   8.1&$ 0.35 \pm 0.40$&$ 0.14 \pm 0.49$&  8.7& 3.9&191.5&-10.99 \\
 29&RXJ1843.4-3314 & 18:43:24.5 & -33:14:31.4 &  11.8&$-0.51 \pm 0.35$&$     \le  0.32$& 14.3& 5.0&194.2&-11.25 \\
 30&RXJ1844.3-3541 & 18:44:19.5 & -35:41:43.5 &  30.0&$ 0.70 \pm 0.19$&$ 0.12 \pm 0.26$& 24.5& 5.8&149.5&-10.92 \\
 31&RXJ1844.5-3723 & 18:44:31.1 & -37:23:59.8 &  43.4&$ 0.62 \pm 0.20$&$ 0.18 \pm 0.26$& 23.2& 5.2&141.1&-10.94 \\
 32&RXJ1844.5-3739 & 18:44:31.3 & -37:39:13.9 &  10.4&$ 0.53 \pm 0.46$&$     \ge  0.24$&  6.9& 3.3&134.9&-10.95 \\
 33&RXJ1844.7-3313 & 18:44:43.4 & -33:13:56.1 &   8.9&$ 0.83 \pm 0.31$&$ 0.78 \pm 0.47$&  7.4& 3.8&206.1&-10.90 \\
 34&RXJ1844.7-3450 & 18:44:45.9 & -34:50:55.2 &   7.9&$ 0.68 \pm 0.65$&$     \ge  0.41$&  4.8& 2.9&185.1&-10.92 \\
 35&RXJ1845.1-3324 & 18:45:10.4 & -33:24:26.3 &  10.6&$-0.28 \pm 0.38$&$ 0.20 \pm 0.62$&  9.6& 3.9&200.1&-11.16 \\
 36&RXJ1845.5-3750 & 18:45:34.4 & -37:50:24.3 &  57.7&$ 0.59 \pm 0.15$&$ 0.28 \pm 0.18$& 39.6& 6.9&143.7&-10.94 \\
 37&RXJ1846.3-3552 & 18:46:18.9 & -35:52:34.0 &   7.9&$     \le -0.34$&$              $&  5.6& 3.0&161.1&-11.52 \\
 38&RXJ1846.7-3604 & 18:46:44.3 & -36:04:40.4 &   8.0&$     \ge -0.09$&$ 0.06 \pm 0.59$&  5.1& 2.9&143.7&-10.87 \\
 39&RXJ1846.7-3636 & 18:46:46.2 & -36:36:21.3 &  29.2&$ 0.07 \pm 0.27$&$ 0.20 \pm 0.36$& 19.1& 4.9&142.0&-11.06 \\
 40&RXJ1846.8-3911 & 18:46:53.3 & -39:11:10.3 &   7.8&$ 0.45 \pm 0.42$&$-0.12 \pm 0.49$&  8.9& 3.6&125.3&-10.97 \\
 41&RXJ1846.9-3503 & 18:46:58.0 & -35:03:59.4 &   9.1&$ 0.84 \pm 0.33$&$     \le -0.59$&  7.9& 3.8&176.3&-10.89 \\
 42&RXJ1847.2-3709 & 18:47:14.3 & -37:09:43.3 &   8.6&$ 0.47 \pm 0.42$&$     \le -0.30$&  6.5& 3.2&136.3&-10.97 \\
 43&RXJ1847.5-3642 & 18:47:32.9 & -36:42:43.9 &   9.6&$ 0.93 \pm 0.44$&$ 0.13 \pm 0.52$&  4.6& 2.3&136.0&-10.88 \\
 44&RXJ1847.7-3606 & 18:47:46.5 & -36:06:12.4 &  29.7&$     \ge  0.68$&$ 0.14 \pm 0.28$& 15.7& 4.4&154.3&-10.87 \\
 45&RXJ1847.7-4024 & 18:47:46.9 & -40:24:14.2 &  69.8&$ 0.00 \pm 0.19$&$ 0.35 \pm 0.24$& 36.7& 6.5&130.2&-11.08 \\
 46&RXJ1847.8-3828 & 18:47:51.9 & -38:28:45.9 &   7.4&$     \ge  0.19$&$     \ge  0.23$&  2.2& 1.7&122.5&-10.87 \\
 47&RXJ1848.2-3415 & 18:48:12.1 & -34:15:41.3 &   8.2&$ 0.42 \pm 0.39$&$-0.16 \pm 0.47$&  9.4& 4.0&202.0&-10.98 \\
 48&RXJ1848.6-3458 & 18:48:36.1 & -34:58:58.0 &   8.0&$ 0.08 \pm 0.59$&$ 0.22 \pm 0.85$&  8.7& 3.8&200.2&-11.06 \\
 49&RXJ1849.0-3734 & 18:49:02.0 & -37:34:22.7 &   9.6&$     \ge  0.40$&$-0.44 \pm 0.41$&  6.0& 3.0&157.8&-10.87 \\
 50&RXJ1849.1-3546 & 18:49:08.7 & -35:46:41.7 &  45.0&$ 0.71 \pm 0.17$&$ 0.56 \pm 0.20$& 22.5& 5.2&193.1&-10.92 \\
 51&RXJ1851.4-3455 & 18:51:26.1 & -34:55:31.6 &   8.1&$     \ge  0.39$&$-0.45 \pm 0.48$&  5.3& 3.0&211.0&-10.87 \\
 52&RXJ1852.1-3607 & 18:52:10.2 & -36:07:24.9 &   7.8&$ 0.19 \pm 0.39$&$-0.57 \pm 0.43$&  9.4& 3.8&171.9&-11.03 \\ 

\end{tabular}
\end{table*}

\begin{table*}
\begin{tabular}{rlccrrrr@{$\pm$}lrr} 
\multicolumn{11}{c}{{\bf Table 2. X-ray data of our sample} (cont.)} \\ \hline
No. & Designation & \multicolumn{2}{c}{$\alpha _{2000}$ (X-ray) $\delta _{2000}$} & $ML$
& HR~1 & HR~2 & \multicolumn{2}{c}{Counts} & \hspace{-.2cm} Exp. [s] & \hspace{-.2cm} $\log ECF$\\ \hline

 53&RXJ1852.3-3700 & 18:52:18.2 & -37:00:20.4 &  36.3&$ 0.90 \pm 0.13$&$ 0.54 \pm 0.25$& 14.7& 4.3&155.5&-10.88 \\
 54&RXJ1852.4-3730 & 18:52:25.6 & -37:30:32.4 &  54.4&$-0.41 \pm 0.19$&$-0.26 \pm 0.38$& 29.4& 6.0&152.8&-11.21 \\
 55&RXJ1853.1-3609 & 18:53:06.5 & -36:09:54.7 &  55.5&$ 0.33 \pm 0.18$&$ 0.40 \pm 0.22$& 37.1& 6.8&190.8&-11.00 \\
 56&RXJ1853.4-4020 & 18:53:26.3 & -40:20:02.8 &   9.2&$-0.18 \pm 0.49$&$     \le  0.01$&  7.7& 3.4&147.6&-11.13 \\
 57&RXJ1853.5-3544 & 18:53:36.0 & -35:44:32.7 &   9.4&$     \ge  0.34$&$ 0.10 \pm 0.48$&  6.4& 3.3&214.1&-10.87 \\
 58&RXJ1854.4-4010 & 18:54:25.7 & -40:10:21.7 &   9.3&$     \ge  0.32$&$ 0.40 \pm 0.48$&  7.1& 3.4&154.8&-10.87 \\
 59&RXJ1854.4-3738 & 18:54:28.2 & -37:38:39.7 &   7.7&$     \ge  0.37$&$ 0.39 \pm 0.44$&  6.0& 3.1&173.8&-10.87 \\
 60&RXJ1854.5-3823 & 18:54:34.3 & -38:23:22.2 &   8.5&$-0.31 \pm 0.42$&$     \le -0.27$& 13.0& 4.6&161.7&-11.18 \\
 61&RXJ1854.9-3600 & 18:54:59.4 & -36:00:26.0 &   8.9&$     \ge  0.40$&$-0.47 \pm 0.44$&  9.5& 4.0&232.8&-10.87 \\
 62&RXJ1855.5-3806 & 18:55:32.5 & -38:06:41.2 &  12.7&$     \ge  0.58$&$ 0.76 \pm 0.32$&  8.4& 4.0&171.9&-10.87 \\
 63&RXJ1856.5-3754 & 18:56:35.1 & -37:54:32.7 &2643.8&$-0.92 \pm 0.02$&$-0.94 \pm 0.16$&637.3&25.5&166.8&-11.46 \\
 64&RXJ1856.6-3545 & 18:56:43.7 & -35:45:23.1 &   8.9&$     \ge  0.07$&$ 0.22 \pm 0.28$&  4.7& 2.7&216.0&-10.87 \\
 65&RXJ1856.7-4021 & 18:56:46.2 & -40:21:10.7 &  12.0&$     \ge  0.52$&$ 0.61 \pm 0.37$&  7.0& 3.5&166.0&-10.87 \\
 66&RXJ1857.3-3635 & 18:57:19.8 & -36:35:49.3 &  10.6&$-0.45 \pm 0.38$&$-0.78 \pm 1.33$& 11.4& 4.2&211.7&-11.23 \\
 67&RXJ1857.3-3554 & 18:57:22.4 & -35:54:19.4 &  10.0&$     \le -0.45$&$     \ge -0.65$&  6.7& 3.3&244.5&-11.52 \\
 68&RXJ1857.5-3732 & 18:57:33.7 & -37:32:57.3 &  10.5&$ 0.78 \pm 0.37$&$ 0.95 \pm 0.37$&  5.8& 2.9&176.8&-10.90 \\
 69&RXJ1857.9-3651 & 18:58:00.1 & -36:51:10.7 &   8.5&$     \le -0.43$&$     \ge -0.83$&  8.3& 3.8&225.4&-11.52 \\
 70&RXJ1858.0-3640 & 18:58:05.3 & -36:40:57.2 &   8.1&$-0.01 \pm 0.41$&$-0.44 \pm 0.47$& 10.3& 4.3&236.1&-11.08 \\
 71&RXJ1858.4-3446 & 18:58:29.4 & -34:46:53.2 &  32.1&$ 0.83 \pm 0.16$&$-0.32 \pm 0.23$& 27.8& 6.4&312.3&-10.90 \\
 72&RXJ1858.6-3630 & 18:58:40.4 & -36:30:38.8 &  12.3&$-0.17 \pm 0.35$&$ 0.42 \pm 0.45$& 12.5& 4.6&257.0&-11.13 \\
 73&RXJ1858.7-3706 & 18:58:43.5 & -37:06:18.3 &  25.1&$-0.19 \pm 0.27$&$ 0.22 \pm 0.49$& 20.9& 5.3&229.4&-11.14 \\
 74&RXJ1859.0-4051 & 18:59:01.1 & -40:51:19.5 &   8.5&$     \ge  0.14$&$     \le -0.23$&  6.0& 3.2&154.6&-10.87 \\
 75&RXJ1859.3-3406 & 18:59:21.7 & -34:06:09.9 &   7.5&$     \ge  0.55$&$-0.16 \pm 0.42$&  6.1& 3.7&300.1&-10.87 \\
 76&RXJ1900.6-3810 & 19:00:38.9 & -38:10:19.5 &  14.2&$ 0.99 \pm 0.15$&$ 0.27 \pm 0.30$& 13.4& 4.9&234.9&-10.87 \\
 77&RXJ1900.8-3453 & 19:00:51.4 & -34:53:08.0 &   7.9&$ 0.99 \pm 0.44$&$ 0.10 \pm 0.39$&  9.9& 4.3&296.8&-10.87 \\
 78&RXJ1900.9-3928 & 19:00:59.6 & -39:28:55.6 &  10.3&$     \ge  0.54$&$-0.43 \pm 0.35$&  8.1& 3.8&190.4&-10.87 \\
 79&RXJ1901.1-3333 & 19:01:08.8 & -33:33:40.8 &   8.9&$     \ge  0.39$&$     \le -0.35$&  3.6& 2.4&233.9&-10.87 \\
 80&RXJ1901.1-3648 & 19:01:09.2 & -36:48:01.7 &   9.8&$ 0.42 \pm 0.39$&$ 0.44 \pm 0.42$&  9.7& 4.0&272.1&-10.98 \\
 81&RXJ1901.4-4022 & 19:01:24.6 & -40:22:31.9 &  11.3&$ 0.63 \pm 0.63$&$-0.17 \pm 0.46$& 11.0& 4.2&188.3&-10.93 \\
 82&RXJ1901.4-3422 & 19:01:26.8 & -34:22:44.8 &  48.4&$ 0.11 \pm 0.19$&$-0.40 \pm 0.23$& 37.7& 7.1&295.0&-11.05 \\
 83&RXJ1901.5-3700 & 19:01:31.4 & -37:00:55.1 &  37.7&$ 0.83 \pm 0.17$&$-0.06 \pm 0.22$& 26.8& 6.1&268.0&-10.90 \\
 84&RXJ1901.6-3652 & 19:01:38.9 & -36:52:55.0 &  38.3&$ 0.80 \pm 0.18$&$ 0.54 \pm 0.20$& 24.6& 5.8&278.8&-10.90 \\
 85&RXJ1901.6-3644 & 19:01:39.0 & -36:44:52.4 &  27.1&$     \ge  0.61$&$ 0.95 \pm 0.15$& 15.2& 4.8&279.1&-10.87 \\
 86&RXJ1902.0-3822 & 19:02:03.1 & -38:22:02.4 &   8.9&$ 0.15 \pm 0.42$&$     \ge  0.21$&  8.3& 3.5&237.5&-11.04 \\
 87&RXJ1902.0-3707 & 19:02:03.1 & -37:07:47.7 &  22.6&$ 0.71 \pm 0.21$&$ 0.58 \pm 0.24$& 17.5& 4.8&268.1&-10.92 \\
 88&RXJ1902.3-3310 & 19:02:20.0 & -33:10:45.8 &   8.0&$-0.82 \pm 0.47$&$              $&  8.1& 3.7&187.4&-11.40 \\
 89&RXJ1902.3-3655 & 19:02:22.7 & -36:55:42.8 &   7.5&$     \ge  0.07$&$ 0.76 \pm 0.38$&  4.8& 2.8&282.5&-10.87 \\
 90&RXJ1902.3-3922 & 19:02:22.8 & -39:22:11.2 &   9.5&$-0.32 \pm 0.44$&$-0.64 \pm 0.85$&  9.4& 3.8&204.3&-11.18 \\
 91&RXJ1902.4-3617 & 19:02:24.0 & -36:17:53.5 &   7.6&$     \ge  0.54$&$-0.06 \pm 0.48$&  4.1& 2.7&306.9&-10.87 \\
 92&RXJ1902.5-3504 & 19:02:31.6 & -35:04:58.8 &   9.5&$     \le -0.58$&$              $&  7.0& 3.6&305.2&-11.52 \\
 93&RXJ1902.7-3418 & 19:02:45.6 & -34:18:36.0 &   8.8&$     \ge  0.47$&$ 0.67 \pm 0.46$&  5.3& 3.3&288.8&-10.87 \\
 94&RXJ1903.0-3906 & 19:03:01.6 & -39:06:09.4 &  12.9&$     \ge  0.50$&$ 0.20 \pm 0.37$& 10.4& 4.1&222.5&-10.87 \\
 95&RXJ1903.0-4009 & 19:03:03.4 & -40:09:09.7 &  18.6&$ 0.63 \pm 0.34$&$ 0.03 \pm 0.34$& 13.6& 4.3&194.3&-10.93 \\
 96&RXJ1904.0-3817 & 19:04:01.7 & -38:17:05.0 &   7.7&$     \ge  0.28$&$     \le -0.39$&  4.4& 2.8&286.9&-10.87 \\
 97&RXJ1904.0-3804 & 19:04:04.0 & -38:04:19.0 &  10.2&$-0.64 \pm 0.39$&$-0.76 \pm 1.61$& 13.5& 4.8&299.3&-11.31 \\
 98&RXJ1904.0-3824 & 19:04:04.7 & -38:24:56.8 &   7.4&$     \le -0.14$&$              $&  5.2& 3.0&293.0&-11.52 \\
 99&RXJ1904.3-3835 & 19:04:23.7 & -38:35:15.9 &  10.6&$     \ge  0.42$&$     \ge  0.36$&  7.4& 3.8&289.8&-10.87 \\
100&RXJ1904.5-3307 & 19:04:34.0 & -33:07:16.0 &   8.0&$ 0.91 \pm 0.38$&$ 0.27 \pm 0.39$&  7.9& 3.7&115.6&-10.88 \\
101&RXJ1904.5-4028 & 19:04:34.4 & -40:28:31.9 &   7.5&$     \le -0.29$&$     \le  0.77$&  7.4& 3.8&218.5&-11.52 \\
102&RXJ1904.6-4048 & 19:04:36.4 & -40:48:24.2 &   8.9&$     \ge  0.42$&$-0.15 \pm 0.44$&  7.3& 3.4&211.3&-10.87 \\
103&RXJ1904.7-3604 & 19:04:43.8 & -36:04:58.5 &  17.8&$ 0.60 \pm 0.26$&$ 0.33 \pm 0.29$& 17.6& 5.1&326.7&-10.94 \\
104&RXJ1905.0-3817 & 19:05:04.9 & -38:17:46.3 &   9.6&$ 0.09 \pm 0.36$&$-0.04 \pm 0.49$& 11.7& 4.3&313.2&-11.06 \\
105&RXJ1905.0-3629 & 19:05:05.2 & -36:29:14.0 &   8.8&$ 0.86 \pm 0.74$&$     \le -0.50$&  6.8& 3.6&332.1&-10.89 \\
106&RXJ1905.3-3855 & 19:05:18.8 & -38:55:15.1 &  12.1&$ 0.70 \pm 0.33$&$ 0.07 \pm 0.35$& 17.5& 5.1&304.2&-10.92 \\
107&RXJ1905.3-4030 & 19:05:19.6 & -40:30:59.1 &   8.5&$-0.34 \pm 0.36$&$-0.28 \pm 0.71$& 12.3& 4.5&247.9&-11.19 \\
108&RXJ1906.4-3703 & 19:06:25.0 & -37:03:34.9 &  25.2&$-0.80 \pm 0.25$&$     \ge -0.10$& 14.3& 4.2&325.1&-11.39 \\
109&RXJ1906.5-3932 & 19:06:31.9 & -39:32:58.1 &   8.8&$-0.93 \pm 0.55$&$     \le  0.06$&  8.6& 3.9&289.6&-11.47 \\

\end{tabular}
\end{table*}

\begin{table*}
\begin{tabular}{rlccrrrr@{$\pm$}lrr} 
\multicolumn{11}{c}{{\bf Table 2. X-ray data of our sample} (cont.)} \\ \hline
No. & Designation & \multicolumn{2}{c}{$\alpha _{2000}$ (X-ray) $\delta _{2000}$} & $ML$
& HR~1 & HR~2 & \multicolumn{2}{c}{Counts} & \hspace{-.2cm} Exp. [s] & \hspace{-.2cm} $\log ECF$\\ \hline

110&RXJ1906.8-3748 & 19:06:51.2 & -37:48:40.3 &  12.8&$     \le -0.55$&$              $&  8.4& 3.4&325.3&-11.52 \\
111&RXJ1907.8-3923 & 19:07:50.8 & -39:23:39.8 &  74.9&$     \ge  0.82$&$ 0.33 \pm 0.18$& 28.8& 5.9&294.9&-10.87 \\
112&RXJ1909.6-3949 & 19:09:40.9 & -39:49:37.6 &  35.1&$ 0.52 \pm 0.22$&$ 0.26 \pm 0.26$& 20.4& 5.0&322.2&-10.96 \\
113&RXJ1909.8-3343 & 19:09:50.9 & -33:43:17.9 &  10.6&$     \le -0.31$&$              $&  7.0& 3.4&222.6&-11.52 \\
114&RXJ1910.8-3854 & 19:10:48.8 & -38:54:48.4 &  20.0&$-0.04 \pm 0.27$&$-0.42 \pm 0.36$& 20.7& 5.4&351.6&-11.09 \\
115&RXJ1911.5-3434 & 19:11:34.8 & -34:34:59.3 &  40.6&$ 0.68 \pm 0.17$&$ 0.13 \pm 0.24$& 26.9& 6.0&323.7&-10.92 \\
116&RXJ1911.7-3641 & 19:11:47.4 & -36:41:54.9 &  11.3&$-0.76 \pm 0.41$&$ 0.81 \pm 1.14$& 10.3& 3.9&320.2&-11.37 \\
117&RXJ1912.6-3431 & 19:12:36.6 & -34:31:40.5 &  16.7&$ 0.57 \pm 0.34$&$ 0.09 \pm 0.36$& 13.8& 4.5&322.2&-10.95 \\
118&RXJ1913.0-3633 & 19:13:01.2 & -36:33:04.4 &  22.2&$ 0.68 \pm 0.24$&$ 0.00 \pm 0.30$& 16.4& 4.7&319.0&-10.92 \\
119&RXJ1913.1-3622 & 19:13:10.2 & -36:22:06.1 &   8.6&$     \ge  0.39$&$-0.59 \pm 0.38$&  7.6& 3.5&318.8&-10.87 \\
120&RXJ1913.3-3530 & 19:13:21.8 & -35:30:40.1 &   8.3&$     \ge  0.27$&$     \le -0.35$&  4.7& 2.8&317.9&-10.87 \\
121&RXJ1913.6-3543 & 19:13:36.6 & -35:43:10.1 &  11.7&$     \ge  0.54$&$     \le -0.39$&  5.8& 3.3&307.7&-10.87 \\
122&RXJ1913.7-3303 & 19:13:46.3 & -33:03:59.8 &   9.3&$ 0.51 \pm 0.32$&$-0.15 \pm 0.42$& 10.8& 4.1&194.9&-10.96 \\
123&RXJ1913.8-3347 & 19:13:50.2 & -33:47:49.1 &  13.9&$-0.06 \pm 0.38$&$ 0.36 \pm 0.55$& 15.8& 5.2&275.4&-11.10 \\
124&RXJ1914.4-3532 & 19:14:28.9 & -35:32:27.2 &   8.1&$     \ge  0.14$&$     \le -0.19$&  7.5& 3.6&292.3&-10.87 \\
125&RXJ1914.6-3601 & 19:14:36.3 & -36:01:12.9 &   8.0&$     \le -0.37$&$              $&  8.4& 4.0&286.1&-11.52 \\
126&RXJ1915.5-3759 & 19:15:32.2 & -37:59:35.3 &   8.5&$     \le -0.49$&$              $&  4.6& 2.7&316.6&-11.52 \\
127&RXJ1915.5-3528 & 19:15:34.3 & -35:28:45.2 &   7.8&$ 0.69 \pm 0.43$&$-0.42 \pm 0.36$&  9.8& 4.2&283.4&-10.92 \\
128&RXJ1915.7-3321 & 19:15:46.7 & -33:21:58.7 &   9.7&$     \ge  0.37$&$-0.59 \pm 0.37$&  5.7& 2.9&228.0&-10.87 \\
129&RXJ1915.9-3350 & 19:15:58.7 & -33:50:47.7 &   9.3&$     \ge  0.31$&$     \le -0.50$&  5.1& 3.2&267.5&-10.87 \\
130&RXJ1916.2-3548 & 19:16:13.1 & -35:48:20.1 &  11.3&$ 0.57 \pm 0.25$&$-0.35 \pm 0.37$& 13.6& 4.6&283.9&-10.94 \\
131&RXJ1916.5-3458 & 19:16:33.7 & -34:58:55.7 &   9.2&$     \le -0.35$&$     \le  0.47$&  9.2& 4.1&292.7&-11.52 \\
132&RXJ1916.9-4024 & 19:16:58.2 & -40:24:32.0 &   7.8&$-0.85 \pm 0.45$&$     \ge -0.50$&  6.0& 2.9&246.5&-11.42 \\
133&RXJ1917.0-3331 & 19:17:04.0 & -33:31:04.4 &  28.0&$     \ge  0.63$&$ 0.77 \pm 0.19$& 17.0& 5.0&219.5&-10.87 \\
134&RXJ1917.2-3931 & 19:17:12.9 & -39:31:25.8 &   7.6&$     \ge  0.20$&$     \le -0.06$&  4.1& 2.5&287.9&-10.87 \\
135&RXJ1917.4-3756 & 19:17:24.3 & -37:56:55.4 & 122.6&$ 0.46 \pm 0.12$&$ 0.08 \pm 0.16$& 58.1& 8.2&310.7&-10.97 \\
136&RXJ1917.8-3927 & 19:17:51.3 & -39:27:33.4 &   7.8&$     \ge  0.41$&$-0.16 \pm 0.47$&  5.7& 3.1&286.2&-10.87 \\
137&RXJ1918.0-4000 & 19:18:00.7 & -40:00:28.6 &   7.6&$ 0.25 \pm 0.46$&$     \ge  0.24$&  6.7& 3.2&250.7&-11.02 \\
138&RXJ1918.2-3823 & 19:18:12.9 & -38:23:05.8 &1044.6&$-0.08 \pm 0.06$&$ 0.20 \pm 0.08$&319.8&18.3&307.7&-11.10 \\
139&RXJ1918.4-3913 & 19:18:25.1 & -39:13:30.9 &   7.7&$ 0.62 \pm 0.37$&$     \le -0.48$&  7.8& 3.6&299.9&-10.94 \\
140&RXJ1919.4-4031 & 19:19:29.1 & -40:31:34.7 &   9.0&$-0.02 \pm 0.40$&$ 0.51 \pm 0.56$& 12.1& 4.4&209.3&-11.09 \\
141&RXJ1919.5-3639 & 19:19:31.5 & -36:39:37.2 &   9.4&$-0.96 \pm 0.51$&$     \le  0.61$&  8.3& 3.9&298.4&-11.49 \\
142&RXJ1919.9-3924 & 19:19:55.3 & -39:24:56.6 &  16.5&$-0.40 \pm 0.31$&$     \ge  0.17$& 14.8& 4.7&306.5&-11.21 \\ 
143&RXJ1919.9-4009 & 19:19:59.1 & -40:09:22.0 &   7.5&$-0.52 \pm 0.36$&$-0.43 \pm 0.82$& 10.4& 4.1&252.7&-11.25 \\
144&RXJ1920.6-3911 & 19:20:41.4 & -39:11:29.6 &   7.9&$ 0.66 \pm 0.36$&$ 0.35 \pm 0.45$&  7.3& 3.3&314.8&-10.93 \\
145&RXJ1920.9-3320 & 19:20:59.0 & -33:20:06.7 &   9.5&$ 0.32 \pm 0.37$&$ 0.12 \pm 0.45$& 10.4& 4.0&257.8&-11.00 \\
146&RXJ1921.1-4002 & 19:21:09.2 & -40:02:14.4 &   9.6&$     \le -0.55$&$              $&  8.6& 4.1&264.9&-11.52 \\
147&RXJ1921.4-3459 & 19:21:28.9 & -34:59:04.4 & 472.1&$-0.10 \pm 0.07$&$-0.04 \pm 0.11$&202.2&14.9&300.4&-11.11 \\
148&RXJ1922.0-3549 & 19:22:01.8 & -35:49:16.1 &   9.8&$     \ge  0.54$&$-0.72 \pm 0.33$&  7.8& 4.0&311.7&-10.87 \\
149&RXJ1922.4-3806 & 19:22:24.3 & -38:06:13.6 &   8.3&$     \ge  0.60$&$-0.06 \pm 0.45$&  6.7& 3.4&303.9&-10.87 \\
150&RXJ1922.9-3545 & 19:22:55.3 & -35:45:39.2 &  21.4&$     \ge  0.68$&$     \ge  0.53$& 10.5& 3.9&303.9&-10.87 \\
151&RXJ1923.3-3658 & 19:23:18.9 & -36:58:39.3 &   9.7&$-0.05 \pm 0.37$&$ 0.30 \pm 0.56$& 10.2& 4.0&320.7&-11.10 \\
152&RXJ1923.6-3551 & 19:23:37.6 & -35:51:39.8 &   8.7&$     \ge  0.41$&$ 0.48 \pm 0.47$&  6.2& 3.2&298.5&-10.87 \\
153&RXJ1923.8-3700 & 19:23:51.9 & -37:00:08.2 &   7.6&$ 0.51 \pm 0.34$&$-0.02 \pm 0.51$&  7.4& 3.3&311.9&-10.96 \\
154&RXJ1923.8-4036 & 19:23:53.3 & -40:36:46.4 &  66.9&$-0.08 \pm 0.19$&$-0.35 \pm 0.27$& 33.1& 6.3&275.6&-11.10 \\
155&RXJ1924.1-3333 & 19:24:09.9 & -33:33:10.2 &   8.0&$ 0.74 \pm 0.40$&$ 0.32 \pm 0.42$&  9.7& 4.2&296.8&-10.91 \\
156&RXJ1924.3-4008 & 19:24:20.2 & -40:08:35.5 &   7.7&$ 0.02 \pm 0.46$&$     \le -0.11$& 10.0& 4.3&309.0&-11.08 \\
157&RXJ1924.4-3959 & 19:24:27.9 & -39:59:29.1 &  10.3&$ 0.53 \pm 0.38$&$ 0.87 \pm 0.32$&  6.9& 3.2&307.9&-10.95 \\
158&RXJ1924.5-3442 & 19:24:34.0 & -34:42:34.6 &  20.5&$-0.29 \pm 0.31$&$ 0.95 \pm 0.47$& 15.0& 4.5&298.9&-11.17 \\
159&RXJ1924.7-3601 & 19:24:47.3 & -36:01:15.1 &   7.5&$-0.09 \pm 0.45$&$ 0.20 \pm 0.77$&  7.7& 3.5&297.7&-11.11 \\
160&RXJ1925.0-3621 & 19:25:02.7 & -36:21:39.1 &  10.1&$ 0.07 \pm 0.40$&$-0.22 \pm 0.54$&  8.2& 3.4&297.3&-11.06 \\
161&RXJ1925.0-3549 & 19:25:03.9 & -35:49:34.4 &   7.6&$     \ge  0.39$&$ 0.60 \pm 0.32$& 10.8& 4.6&297.0&-10.87 \\
162&RXJ1925.3-3413 & 19:25:21.4 & -34:13:06.4 &   7.4&$     \le -0.49$&$     \le  0.88$&  9.6& 4.3&312.5&-11.52 \\
163&RXJ1926.5-3426 & 19:26:31.8 & -34:26:25.7 &   7.6&$     \le -0.47$&$     \ge -0.60$&  6.8& 3.5&300.9&-11.52 \\
164&RXJ1927.1-3713 & 19:27:06.2 & -37:13:42.2 &  14.6&$     \ge  0.65$&$ 0.77 \pm 0.30$& 10.2& 4.0&316.9&-10.87 \\
165&RXJ1927.4-3847 & 19:27:28.0 & -38:47:06.9 &  28.3&$     \ge  0.71$&$ 0.29 \pm 0.25$& 17.3& 5.0&307.7&-10.87 \\
166&RXJ1927.9-3955 & 19:27:55.0 & -39:55:03.8 &   7.9&$     \le -0.44$&$     \le  0.98$&  5.5& 3.0&325.6&-11.52 \\

\end{tabular}
\end{table*}

\begin{table*}
\begin{tabular}{rlccrrrr@{$\pm$}lrr} 
\multicolumn{11}{c}{{\bf Table 2. X-ray data of our sample } (cont.)} \\ \hline
No. & Designation & \multicolumn{2}{c}{$\alpha _{2000}$ (X-ray) $\delta _{2000}$} & $ML$
& HR~1 & HR~2 & \multicolumn{2}{c}{Counts} & \hspace{-.2cm} Exp. [s] & \hspace{-.2cm} $\log ECF$\\ \hline

167&RXJ1928.0-4050 & 19:28:04.8 & -40:50:04.0 &  46.2&$ 0.60 \pm 0.17$&$ 0.00 \pm 0.22$& 31.1& 6.5&322.3&-10.94 \\
168&RXJ1928.2-3352 & 19:28:13.1 & -33:52:16.1 &  11.9&$     \ge  0.45$&$ 0.03 \pm 0.45$&  8.2& 3.5&281.6&-10.87 \\
169&RXJ1928.2-3915 & 19:28:13.3 & -39:15:13.5 &   7.4&$ 0.76 \pm 0.70$&$     \le -0.33$&  5.1& 3.0&320.8&-10.91 \\
170&RXJ1928.3-4031 & 19:28:20.6 & -40:31:19.7 &  10.3&$     \ge  0.48$&$ 0.85 \pm 0.54$&  9.4& 4.4&325.7&-10.87 \\
171&RXJ1928.5-3508 & 19:28:31.6 & -35:08:02.0 & 268.9&$ 0.83 \pm 0.07$&$ 0.27 \pm 0.11$& 88.8& 9.9&307.2&-10.90 \\
172&RXJ1928.7-3646 & 19:28:46.3 & -36:46:38.6 &  13.6&$     \ge  0.52$&$ 0.15 \pm 0.35$&  9.7& 3.7&308.9&-10.87 \\
173&RXJ1929.0-3805 & 19:29:03.2 & -38:05:15.8 &   8.6&$ 0.39 \pm 0.57$&$     \ge  0.42$&  7.5& 4.0&324.6&-10.98 \\
174&RXJ1929.2-4055 & 19:29:16.0 & -40:55:20.4 &   9.1&$     \le -0.43$&$              $&  5.4& 2.8&329.9&-11.52 \\
175&RXJ1929.6-4018 & 19:29:36.5 & -40:18:53.6 &  39.3&$ 0.80 \pm 0.17$&$ 0.07 \pm 0.23$& 26.9& 5.9&344.0&-10.90 \\
176&RXJ1929.8-3702 & 19:29:50.0 & -37:02:35.7 &  14.8&$-0.74 \pm 0.35$&$     \ge -0.03$& 12.9& 4.6&318.8&-11.36 \\
177&RXJ1929.9-3338 & 19:29:57.4 & -33:38:16.7 &   8.9&$     \ge  0.38$&$-0.21 \pm 0.43$&  8.5& 3.9&281.0&-10.87 \\
178&RXJ1930.0-3302 & 19:30:03.1 & -33:02:55.2 &   7.9&$     \le -0.57$&$              $&  6.0& 3.7&248.9&-11.52 \\
179&RXJ1930.6-3525 & 19:30:39.2 & -35:25:47.5 &  13.8&$     \ge  0.61$&$-0.19 \pm 0.31$& 13.6& 4.7&324.4&-10.87 \\
180&RXJ1931.1-3441 & 19:31:08.1 & -34:41:06.6 &  11.0&$     \ge  0.39$&$     \le -0.47$&  9.6& 4.1&319.5&-10.87 \\
181&RXJ1931.4-4008 & 19:31:25.1 & -40:08:46.9 &   8.8&$ 0.65 \pm 0.39$&$ 0.72 \pm 0.38$&  6.4& 3.2&344.1&-10.93 \\
182&RXJ1931.5-3340 & 19:31:32.5 & -33:40:15.9 &   7.8&$     \le -0.51$&$              $&  5.8& 3.3&272.9&-11.52 \\
183&RXJ1931.6-3354 & 19:31:37.7 & -33:54:38.7 &  96.7&$ 0.93 \pm 0.07$&$ 0.76 \pm 0.10$& 52.5& 8.4&285.2&-10.88 \\
184&RXJ1932.3-3804 & 19:32:19.3 & -38:04:58.6 &   7.4&$     \ge  0.41$&$ 0.67 \pm 0.32$&  6.0& 3.3&324.9&-10.87 \\
185&RXJ1932.5-3506 & 19:32:33.5 & -35:06:32.9 &   7.8&$     \ge  0.47$&$-0.46 \pm 0.40$&  8.0& 3.8&313.0&-10.87 \\
186&RXJ1932.5-3837 & 19:32:34.1 & -38:37:00.5 &  10.6&$     \ge  0.22$&$     \le -0.54$&  6.5& 3.6&329.6&-10.87 \\
187&RXJ1932.7-3432 & 19:32:44.3 & -34:32:22.7 &  11.8&$-0.34 \pm 0.31$&$ 0.50 \pm 0.63$& 13.4& 4.6&305.7&-11.19 \\
188&RXJ1932.7-3858 & 19:32:45.6 & -38:58:39.6 &   9.5&$ 0.88 \pm 0.54$&$     \le -0.45$&  7.4& 3.6&336.0&-10.89 \\
189&RXJ1932.9-3554 & 19:32:56.0 & -35:54:23.0 &   7.9&$     \ge  0.30$&$     \le -0.13$&  5.6& 3.2&314.9&-10.87 \\
190&RXJ1933.2-3337 & 19:33:18.1 & -33:37:29.4 &  10.5&$ 0.55 \pm 0.50$&$     \ge  0.26$&  8.8& 3.9&284.4&-10.95 \\
191&RXJ1933.3-3812 & 19:33:20.2 & -38:12:30.8 &  14.7&$ 0.68 \pm 0.40$&$-0.08 \pm 0.39$& 10.9& 4.0&350.1&-10.92 \\
192&RXJ1933.5-3743 & 19:33:33.5 & -37:43:46.7 &   7.4&$     \le -0.58$&$              $&  5.6& 3.5&336.1&-11.52 \\
193&RXJ1933.6-3453 & 19:33:41.6 & -34:53:27.1 &  11.8&$ 0.77 \pm 0.34$&$ 0.48 \pm 0.37$& 13.2& 4.7&321.7&-10.91 \\
194&RXJ1933.8-3522 & 19:33:51.3 & -35:22:53.1 &   7.6&$-0.29 \pm 0.47$&$     \le -0.03$&  9.0& 3.9&320.6&-11.17 \\
195&RXJ1934.5-3620 & 19:34:33.6 & -36:20:43.0 &   9.0&$     \ge  0.51$&$ 0.06 \pm 0.42$&  6.8& 3.7&318.2&-10.87 \\
196&RXJ1934.6-3433 & 19:34:41.1 & -34:33:33.2 &   7.7&$     \le -0.40$&$              $&  9.0& 4.2&328.1&-11.52 \\
197&RXJ1934.7-3805 & 19:34:46.7 & -38:05:06.0 &  52.9&$ 0.76 \pm 0.18$&$ 0.26 \pm 0.20$& 30.8& 6.3&342.6&-10.91 \\
198&RXJ1935.5-3536 & 19:35:34.2 & -35:36:10.9 &   9.9&$ 0.42 \pm 0.47$&$ 0.87 \pm 0.57$&  8.3& 3.5&329.8&-10.98 \\
199&RXJ1936.0-3325 & 19:36:03.6 & -33:25:59.1 &   8.8&$ 0.89 \pm 0.38$&$     \ge  0.16$&  4.5& 2.7&274.4&-10.88 \\
200&RXJ1936.0-4002 & 19:36:05.6 & -40:02:24.6 &  10.7&$-0.21 \pm 0.39$&$     \ge  0.24$&  9.9& 3.8&335.3&-11.14 \\
201&RXJ1936.9-3429 & 19:36:56.9 & -34:29:28.0 &   9.4&$ 0.28 \pm 0.42$&$     \le -0.44$&  8.7& 3.8&305.6&-11.01 \\
202&RXJ1937.2-3958 & 19:37:12.6 & -39:58:01.1 &  22.1&$ 0.41 \pm 0.26$&$ 0.21 \pm 0.29$& 21.6& 5.6&333.6&-10.98 \\
203&RXJ1937.2-4010 & 19:37:13.2 & -40:10:21.5 &  11.8&$     \ge  0.46$&$ 0.08 \pm 0.44$&  8.8& 4.0&335.8&-10.87 \\
204&RXJ1938.0-3801 & 19:38:03.0 & -38:01:05.8 &  10.3&$ 0.04 \pm 0.49$&$     \ge  0.46$&  7.7& 3.6&346.1&-11.07 \\
205&RXJ1938.1-3914 & 19:38:09.4 & -39:14:02.4 &   9.3&$     \ge  0.36$&$     \ge  0.42$&  3.6& 2.6&345.0&-10.87 \\
206&RXJ1938.2-4006 & 19:38:13.1 & -40:06:38.1 &   7.8&$ 0.55 \pm 0.40$&$     \ge  0.50$&  7.6& 4.0&333.7&-10.95\\ \hline

\end{tabular}

\end{table*}

In Table 2, we list all RASS X-ray sources detected in our study area
with X-ray position, likelihood for existence, hardness ratio, 
number of counts, exposure time, and individual energy conversion factor
\begin{equation}
\label{ecf}
ECF = \left( 5.30 \cdot HR~1 + 8.31 \right) \cdot 10 ^{-12}~erg~cm ^{-2}~cts ^{-1}
\end{equation}
according to Schmitt et al. (1995).

In Table 3, we list the optical counterparts to the X-ray sources, most of
which are the closest counterparts, but in four cases, the most likely 
counterpart turned out to be not the closest potential optical counterpart.
Table 3 gives optical positions, offsets between X-ray and optical positions,
V magnitudes, X-ray to optical flux ratios, identification information,
e.g. whether it is a new young star, and remarks, e.g. on previously known 
stars found among our X-ray sources.
The typical positional error of RASS sources is 40 arc sec (Neuh\"auser et al. 1995), 
which we allow as offset between optical and X-ray position for source identification.
We searched for stellar counterparts in Simbad, the Hubble Space Telescope
Guide Star Catalog (GSC), and the NASA Extragalactic Database (NED).
We list in Table 3 the optical positions, V magnitude, X-ray to optical flux
ratio, as well as some more information on the identified counterparts.
See Sect. \ref{ocp} for details on the identification.
For calculating the X-ray to optical flux ratio, we use
\begin{equation}
\label{fv}
f_{V} = 10 ^{ -0.4 \cdot (V + 13.42) }~erg~cm ^{-2}~s ^{-1}
\end{equation}
(with $V$ in $mag$) and the X-ray flux, which we get from the data 
in Table 2 via
\begin{equation}
\label{fx}
f_{X} = ECF \cdot counts~/~exposure
\end{equation}
For sources where we can list only an upper or lower limit to HR~1
(in Table 2), we use HR~$1 = 1$ or HR$~1 = -1$, respectively,
for calculating the flux according to Eq. (\ref{ecf}).
The X-ray to optical flux ratio already gives some hints on
the nature of an unidentified X-ray source: 
Extragalactic X-ray sources are optically faint but X-ray bright
with $\log (f_{X}/f_{V})~\ge~-0.5$, while normal stars usually have 
$\log (f_{X}/f_{V})~\le~-1$ 
(Stocke et al. 1983, Motch et al. 1998).

\begin{table*}
\label{tab3ocp}
\caption []{ {\bf Optical counterparts.} \\
For each X-ray source listed in Table 2, we list here the optical 
counterpart from Simbad, GSC, or NED, unless no counterpart is 
found within $40$ arc sec around the X-ray position
(in all but four cases, these counterparts are the closest ones).
We list running number as in Table 2, optical position, 
offset between X-ray and optical position (in arc seconds),
optical magnitude ($V$ in mag), and the (log of the) X-ray to
optical flux ratio. 
In the column ID, we list the nature of the counterpart, i.e. 
{\bf y} for new {\bf young} star listed in Table 4,            
{\bf z} for {\bf zero-age} main-sequence stars (i.e. low lithium),
{\bf d} for {\bf dKe} or {\bf dMe} stars,
{\bf e} for {\bf extra-galactic},
{\bf n} for {\bf neither} of the above (mostly old stars),
{\bf p} for {\bf previously} known TTS or non-TTS (not observed
optically except TY CrA); counterparts without any entry in the column ID 
have not been observed optically.
Finally, some remarks are given as found in Simbad (like proper motion 
PM as $[\mu _{\alpha} \cdot \cos \delta,\mu _{\delta}]$ 
in milli arc seconds per year and radial velocity RV in $km~s^{-1}$). 
or from our optical follow-up observations.
Data on optical counterparts are taken from Simbad or NED,
if remarks are listed, otherwise from GSC.
Some GSC counterparts appear on several GSC plates which may have different colors
and different filters; we have always chosen the closest counterpart.
For stars with V mag given with colons, positions and magnitudes are estimated 
from the DSS charts, because the identified counterpart is different from the
relevant Simbad/GSC/NED counterpart.
}

\begin{tabular}{rccrrccl} \hline
No. & \multicolumn{2}{c}{Optical position} & $\Delta$ & $V$ & $\log$ & ID & Remarks \\ 
Tab. 2 & $\alpha _{2000}$ & $\delta _{2000}$ & ["] & mag & $f_{X}/f_{V}$ & & \\ \hline

  1 & 18:35:06.2 & -34:04:02.8 & 28 & 13.9 &$ -1.15 $&    & \\
  3 & 18:35:46.6 & -32:59:31.2 & 12 &  9.8 &$ -1.24 $&  p & globular cluster NGC 6652 \\
  8 & 18:36:39.5 & -34:51:25.0 & 16 & 13.0 &$ -1.45 $&  y & \\
  9 & 18:37:17.6 & -34:42:42.2 &  8 & 11.3 &$ -2.31 $&  n & \\
 11 & 18:38:20.2 & -35:23:37.2 &  5 & 12.2 &$ -2.26 $&  d & dKe star \\
 12 & 18:39:04.9 & -37:05:27.8 & 22 & 13.1 &$ -1.30 $&  n & \\
 13 & 18:39:05.3 & -37:26:21.8 & 16 & 10.9 &$ -2.35 $&  y & \\
 17 & 18:40:37.6 & -37:28:18.1 & 28 & 11.2 &$ -2.27 $&  d & dKe star \\
 19 & 18:40:53.3 & -35:46:44.6 & 22 & 14.5:&$ -0.84 $&  y & \\
 23 & 18:41:48.6 & -35:25:43.6 & 14 &  9.7 &$ -2.44 $&  y & \\
 27 & 18:42:58.0 & -35:32:42.9 & 10 & 12.2 &$ -1.73 $&  y & \\
 30 & 18:44:21.9 & -35:41:43.6 & 29 & 11.3 &$ -1.81 $&  y & \\
 31 & 18:44:31.1 & -37:23:34.3 & 26 & 13.1 &$ -1.13 $&  y & \\
 35 & 18:45:09.3 & -33:24:03.9 & 26 & 12.3 &$ -2.18 $&  n & \\
 36 & 18:45:34.8 & -37:50:19.6 &  7 &  9.2 &$ -2.45 $&  y & HD 173148, G5 V, PM=[4.7,-32.0] \\
 38 & 18:46:43.9 & -36:04:52.2 & 13 & 12.0 &$ -2.14 $&    & IRAS 18433-3608 \\
 39 & 18:46:45.6 & -36:36:18.1 &  8 & 10.3 &$ -2.45 $&  y & \\
 42 & 18:47:14.0 & -37:09:48.3 &  6 & 12.1 &$ -2.09 $&    & \\
 45 & 18:47:44.6 & -40:24:22.2 & 34 &  5.2 &$ -4.18 $&  n & $\mu$ CrA, G5.5 I, PM=[24.6,-18.6], RV=-18.2, d=120~pc \\
 48 & 18:48:35.8 & -34:58:20.4 & 38 & 13.3 &$ -1.73 $&    & \\
 53 & 18:52:17.3 & -37:00:12.0 & 14 & 13.9 &$ -0.98 $&  y & IRAS 18489-3703 \\
 54 & 18:52:24.8 & -37:30:35.6 & 10 & 12.5 &$ -1.57 $&  d & dMe star \\
 55 & 18:53:06.0 & -36:10:22.8 & 28 &  9.6 &$ -2.47 $&  y & HD 174656, G6 IV, PM=[-1.2,-34.0] \\
 59 & 18:54:29.0 & -37:39:04.5 & 26 & 11.8 &$ -2.23 $&    & \\
 63 & 18:56:37.3 & -37:54:26.9 & 26 & 25.7 &$ +4.77 $&  p & neutron star RXJ1856.5-3754 \\
 64 & 18:56:44.0 & -35:45:31.9 & 10 & 13.0 &$ -2.08 $&  y & \\
 65 & 18:56:49.2 & -40:21:07.0 & 34 & 14.9 &$ -0.92 $&    & \\
 68 & 18:57:34.1 & -37:32:32.3 & 25 & 15.0:&$ -1.02 $&  y & \\
 73 & 18:58:43.4 & -37:06:26.5 &  7 &  4.9 &$ -4.86 $&  p & $\epsilon$ CrA, F2 V, W~UMa-type, d=30~pc \\
 77 & 19:00:49.5 & -34:52:49.2 & 33 &  8.4 &$ -3.61 $&    & HD 176247, G1 V, PM=[23.4,-24.0] \\
 80 & 19:01:09.5 & -36:47:51.7 & 11 & 12.7 &$ -1.97 $&  y & VSS VIII-27 \\
 81 & 19:01:26.7 & -40:22:34.0 & 24 & 13.6 &$ -1.37 $&  n & \\
 82 & 19:01:28.7 & -34:22:35.5 & 25 &  8.2 &$ -3.29 $&  y & HD 176383, F5 V, PM=[9.9,-46.8] \\
 83 & 19:01:34.9 & -37:00:55.8 &  7 & 11.3 &$ -2.03 $&  p & CrAPMS 1, K1 IV, wTTS \\
 84 & 19:01:40.8 & -36:52:34.2 & 30 &  9.5 &$ -2.79 $&  p & TY CrA, Herbig Ae/Be PMS star \\
 85 & 19:01:40.5 & -36:44:31.9 & 27 & 13.0:&$ -1.56 $&  y & VSS VIII-26 \\
 87 & 19:02:01.9 & -37:07:43.2 & 14 & 10.4 &$ -2.57 $&  p & CrAPMS 2, G5 IV, wTTS, PM=[0.0,-33.0] \\
 89 & 19:02:22.1 & -36:55:40.8 &  2 & 13.8 &$ -1.60 $&  p & CrAPMS 3, K2 IV, wTTS \\
 90 & 19:02:22.7 & -39:22:21.9 & 11 & 14.2 &$ -1.46 $&    & \\
 91 & 19:02:25.9 & -36:17:39.0 & 32 &      &$       $&  e & galaxy CGMW 4-4634 \\
 93 & 19:02:43.6 & -34:19:00.4 & 35 & 13.7 &$ -1.75 $&    & \\
 95 & 19:03:00.6 & -40:09:16.8 & 33 & 13.4 &$ -1.35 $&  n & \\
 97 & 19:03:58.4 & -38:04:01.0 & 70 & 13.1 &$ -2.04 $&  z & \\
102 & 19:04:38.8 & -40:48:15.4 & 29 & 11.4 &$ -2.39 $&    & \\
108 & 19:06:24.8 & -37:03:41.7 &  7 &  5.0 &$ -5.37 $&  n & $\gamma$ CrA, F8 V, PM=[95,-274], RV=-51.0, d=21~pc \\
110 & 19:06:52.5 & -37:48:37.6 & 16 &  6.2 &$ -5.26 $&  n & HR 7232, G5 IV, d=17~pc \\
111 & 19:07:50.4 & -39:23:32.2 &  9 & 14.2 &$ -0.84 $&  z & \\

\end{tabular}

\end{table*}

\newpage

\begin{table*}

\begin{tabular}{rccrrccl} 
\multicolumn{8}{c}{ {\bf Table 3. Closest optical counterparts} (cont.) } \\ \hline
No. & \multicolumn{2}{c}{Optical position} & $\Delta$ & $V$ & $\log$ & ID & Remarks \\ 
 & $\alpha _{2000}$ & $\delta _{2000}$ & ["] & mag & $f_{X}/f_{V}$ & & \\ \hline

112 & 19:09:39.8 & -39:49:38.4 & 16 &  6.5 &$ -4.19 $&  n & HR 7255, K1 III, PM=[-4.9,-21.7], d=107~pc \\
114 & 19:10:47.8 & -38:54:34.7 & 19 &  7.6 &$ -3.92 $&  n & HD 178558, F5 V, PM=[26.7,-33.5], d=61~pc \\
115 & 19:11:34.7 & -34:35:09.1 & 10 & 10.7 &$ -2.37 $&  d & dKe star \\
116 & 19:11:47.1 & -36:41:42.9 &  7 & 13.5 &$ -2.09 $&  d & dMe star \\
117 & 19:12:35.8 & -34:31:31.8 & 14 & 11.4 &$ -2.38 $&  n & \\
119 & 19:13:10.5 & -36:21:46.0 & 20 & 12.9 &$ -1.98 $&    & \\
122 & 19:13:44.8 & -33:04:06.6 & 20 & 14.1 &$ -1.23 $&    & \\
123 & 19:13:51.8 & -33:48:21.4 & 38 & 14.6 &$ -1.13 $&    & \\
127 & 19:15:32.5 & -35:28:49.2 & 23 & 14.4 &$ -1.23 $&    & \\
128 & 19:15:46.7 & -33:22:06.3 &  7 & 12.3 &$ -2.17 $&  z & see (1) \\
130 & 19:16:13.8 & -35:48:11.7 & 11 & 14.0 &$ -1.28 $&  d & dKe star \\
135 & 19:17:23.8 & -37:56:50.4 &  8 &  9.9 &$ -2.37 $&  y & SAO 211129, K2, PM=[3.5,-31.0] \\
138 & 19:18:12.4 & -38:23:04.2 &  5 &  8.6 &$ -2.28 $&  n & HD 180445, G8 V, PM=[99.5,-93.0], d=42~pc \\
139 & 19:18:27.0 & -39:13:01.4 & 37 & 13.8 &$ -1.64 $&    & \\
141 & 19:19:31.0 & -36:39:30.7 & 12 &  7.2 &$ -4.80 $&  n & HD 180802, F7 V, PM=[36.4,-95.4], d=49~pc \\
142 & 19:19:54.7 & -39:25:10.1 & 13 &  9.2 &$ -3.34 $&  n & HD 180863, G8/K0 III, PM=[10.4,-126.0] \\
145 & 19:21:01.0 & -33:20:28.7 & 34 & 14.1 &$ -1.38 $&    & \\
147 & 19:21:29.7 & -34:59:00.6 & 11 &  6.5 &$ -3.32 $&  z & HR 7330, G1.5 V, d=21~pc, see (2) \\
151 & 19:23:20.0 & -36:58:31.0 & 16 & 11.0 &$ -2.83 $&    & \\
154 & 19:23:53.0 & -40:36:56.5 & 10 &  4.0 &$ -5.05 $&  p & $\alpha$ Sgr, B8 V, PM=[32.7,-120.8], RV=-0.7, d=52~pc \\
155 & 19:24:07.5 & -33:33:30.0 & 36 & 15.6 &$ -0.81 $&    & \\
158 & 19:24:34.9 & -34:42:37.9 & 11 & 15.3 &$ -0.97 $&  d & dMe star \\
165 & 19:27:26.7 & -38:46:39.4 & 32 & 15.6 &$ -0.51 $&  n & \\
166 & 19:27:56.6 & -39:54:39.4 & 30 & 13.1 &$ -2.66 $&    & \\
167 & 19:28:05.5 & -40:50:04.1 &  7 &  8.2 &$ -3.31 $&  p & HD 182776, K2.5 III, RS~CVn-type, d=240~pc \\ 
169 & 19:28:12.4 & -39:14:56.5 & 20 & 15.3 &$ -1.24 $&    & \\
171 & 19:28:31.9 & -35:07:58.8 &  3 &  8.7 &$ -2.60 $&  z & HD 182928, G5 V, PM=[-16.3,-14.4], d=234~pc, see (1) \\
172 & 19:28:48.7 & -36:46:19.5 & 35 & 12.5 &$ -2.01 $&  n & \\
174 & 19:29:13.1 & -40:55:15.0 & 34 & 13.7 &$ -2.47 $&  n & \\
175 & 19:29:35.0 & -40:18:39.2 & 22 & 12.9 &$ -1.48 $&  z & IRAS 19261-4024, see (1) \\
176 & 19:29:51.1 & -37:02:16.7 & 21 &  9.1 &$ -3.74 $&  n & HD 183198, G6 V, PM=[-22.2,-14.7], d=46~pc \\
177 & 19:29:57.8 & -33:38:16.6 &  5 & 13.1 &$ -1.76 $&    & \\
178 & 19:30:01.6 & -33:02:43.1 & 22 & 12.6 &$ -2.73 $&    & \\
179 & 19:30:38.3 & -35:26:21.8 & 36 & 13.4 &$ -1.54 $&  n & \\
181 & 19:31:22.9 & -40:08:19.8 & 37 & 12.2 &$ -2.40 $&  n & \\
183 & 19:31:38.7 & -33:54:43.2 & 16 & 17.0 &$ +0.56 $&  e & galaxy PKS 1928-340 \\
187 & 19:32:43.8 & -34:32:14.4 & 10 & 14.1 &$ -1.52 $&  n & \\
189 & 19:32:54.1 & -35:54:33.7 & 25 & 14.9 &$ -1.28 $&    & \\
190 & 19:33:17.8 & -33:37:43.0 & 14 & 15.4 &$ -0.93 $&  n & \\
191 & 19:33:19.6 & -38:12:13.6 & 19 & 12.0:&$ -2.26 $&  z & \\
193 & 19:33:40.4 & -34:53:32.0 & 15 & 12.5 &$ -1.92 $&    & \\
195 & 19:34:32.6 & -36:21:11.6 & 31 & 13.4 &$ -1.80 $&  z & \\
197 & 19:34:46.6 & -38:05:15.2 &  9 &  8.9 &$ -3.03 $&  n & HD 184189, M2 III, 19" binary \\
199 & 19:36:01.5 & -33:25:42.9 & 31 & 14.8 &$ -1.38 $&    & \\
200 & 19:36:04.3 & -40:02:58.6 & 37 & 12.6 &$ -2.27 $&  z & \\
202 & 19:37:16.3 & -39:58:01.4 & 56 & 19.0 &$  0.80 $&  e & quasar PKS 1933-400 \\ \hline

\end{tabular}

Notes:
(1) Maybe weak lithium (or noise), complex H$\alpha$, maybe double-lined, 
could be ZAMS, RS~CVn-type, dKe/dMe, or a cool Algol. 
(2) Our detection of weak lithium with 
$W({\lambda}$(Li)$= 0.19 \simeq W_{\lambda}$(Ca) 
confirms the classification of HR 7330 as member of the $\sim$~200 Myr
old nearby Castor moving group by Barrado y Navascues (1998).

\end{table*}

\section {Optical follow-up spectroscopy }
\label{ocp}

Out of the 206 X-ray sources, 89 have one or several nearby ($\le 40$ arc sec) potential 
optical counterparts (brighter than $V=16$ mag) in Simbad and/or GSC, including four 
previously known PMS stars (namely CrAPMS 1, 2, 3, and TY CrA)
and four previously known older stars. In addition, there is one optically faint neutron 
star ($V=25.7$ mag, see footnote 2) and three extra-galactic sources (see Table 3). 
To identify a large fraction of the remaining 81 unidentified X-ray sources,
we performed low-resolution spectroscopy of 148 stars, 
which are potential optical counterparts to 56 of those 81 X-ray sources.  

Low-dispersion spectra were taken with the Boller \& Chivens spectrograph of the
ESO 1.5m telescope on La Silla in twelve nights, namely 1995 July 16/17 to 21/22 
and 1996 July 20/21 to 25/26. The wavelength range is 4500 to 6850\AA~and
the spectral resolution is $\sim 2.5$\AA . For technical details and data 
reduction, we refer to the relevant part of Sect. 2.2 in Walter et al. (1997).

In addition to those potential RASS source counterparts, we also observed
a number of previously known or suspected young stars in CrA, because their lithium
line strength was not known, see Table 1 for those data;
most of these spectra were taken with the Boller \& Chivens spectrograph.
However, the stars HBC 673, 675, 677, and CrAPMS 3/c were observed with 
the ESO-3.5m-NTT\footnote{taken during ESO program 63.L-0023 at the end of the 
night 18/19 April 1999, when the main targets of that run were not visible anymore}. 
Here, we observed in the red medium-dispersion mode (EMMI red arm, 
CCD $\#~36$, grating 6) at a resolution of 5500 in the wavelength range 
from 6160 to 7740 \AA . While we could not detect lithium in HBC 673 and 675, 
we confirmed HBC 677 to be a cTTS and CrAPMS 3/c to be a wTTS (see Table 1).

For most of the RASS counterparts with detected lithium (in low-resolution spectra), 
we then took high-resolution spectra to confirm their youth.
The high dispersion spectra were obtained with the CTIO 4m telescope and
echelle spectrograph on 14 to 17 July 1998.
We used the 226-3 cross disperser and the 31.6 l/mm
echelle with the red optics. We used the SITe 2K \#6 CCD detector, at a gain
of 5, corresponding to about 1~e$^-$ per ADU and a read noise of about
3~e$^-$. We used the GG385 filter for order sorting. We used a 150~$\mu$m
(1~arcsec) slit and decker \#9 (3.3~arcsec) for the stellar observations.
The seeing was generally 1 to 1.5~arcsec, and there were some clouds on 3 of the
4 nights. The spectra cover the range from roughly 4400\AA\ through 7500\AA~at 
a resolution of 25000.
We obtained projector flat images to flatten the spectra. A Th-Ar
comparison source was observed before and after each telescope slew.
Each stellar observation was made in three parts to facilitate cosmic ray
removal.  Initial reductions were undertaken at CTIO, using the IRAF DOECSLIT
package. We corrected for bias, extracted the orders, divided by the flats,
and solved for the
dispersion. The data were rebinned to a linear wavelength scale in each order.
We removed the global background (the scattered light
correction) but did not attempt to subtract the local background.
The data were further reduced using IDL.
We flattened the spectra in each order to remove any residual curvature
left from the original flat division. We trimmed the ends
of the orders.
We then filtered the three individual spectra of each object
to remove cosmic rays, and coadded the spectra.
We determine radial velocities by cross-correlating the spectra against
the sky spectra. We expect an internal precision of about 1~km/s, except
for the targets with poorer S/N.

\section{Results of the spectroscopy}
\label{resspec}

Stars which fullfill {\em all}~the following conditions (according to their high-resolution
spectra, if available, otherwise we use the low-resolution spectra) are classified 
as new PMS stars, i.e. appear as such in Table 4:
\begin{itemize}
\item Spectral type later than mid F, {\em and}
\item lithium 6708\AA~line with $W_{\lambda}$(Li)$~\ge~0.1$\AA , {\em and}
\item more Li than ZAMS stars of the same spectral type, {\em and}
\item $W_{\lambda}$(Li)$~\ge~W_{\lambda}$(Ca) 
(for spectral types F, G, and K)\footnote{X-ray emitting M-type stars with lithium 
are always pre-MS stars, as they burn all Lithium before reaching the MS
(see e.g. Covino et al. 1997).}.
\end{itemize}
All these stars are classified as {\bf young} stars 
(letter {\bf y} in column ID in Table 3). 

Table 4 lists for all newly discovered TTS their official designation,
PMS type (i.e. whether wTTS or cTTS), the spectral type, thje H$\alpha$
and lithium equivalent widths, the projected rotational velocity, the 
radial velocity, the X-ray luminosity, and some remarks, e.g. on binarity.

All stars with a detected lithium line, which is weaker than $0.1$\AA , 
and all F-, G-, and K-type stars with a detected lithium line, 
which is weaker than the Ca 6718\AA~line, are classified as {\bf ZAMS} stars 
(letter {\bf z} in column ID in Table 3).
Stars without detectable lithium, but H$\alpha$ emission are classified as 
{\bf dKe or dMe} stars (letter {\bf d} in column ID in Table 3).

\begin{figure*}
\vbox{\psfig{figure=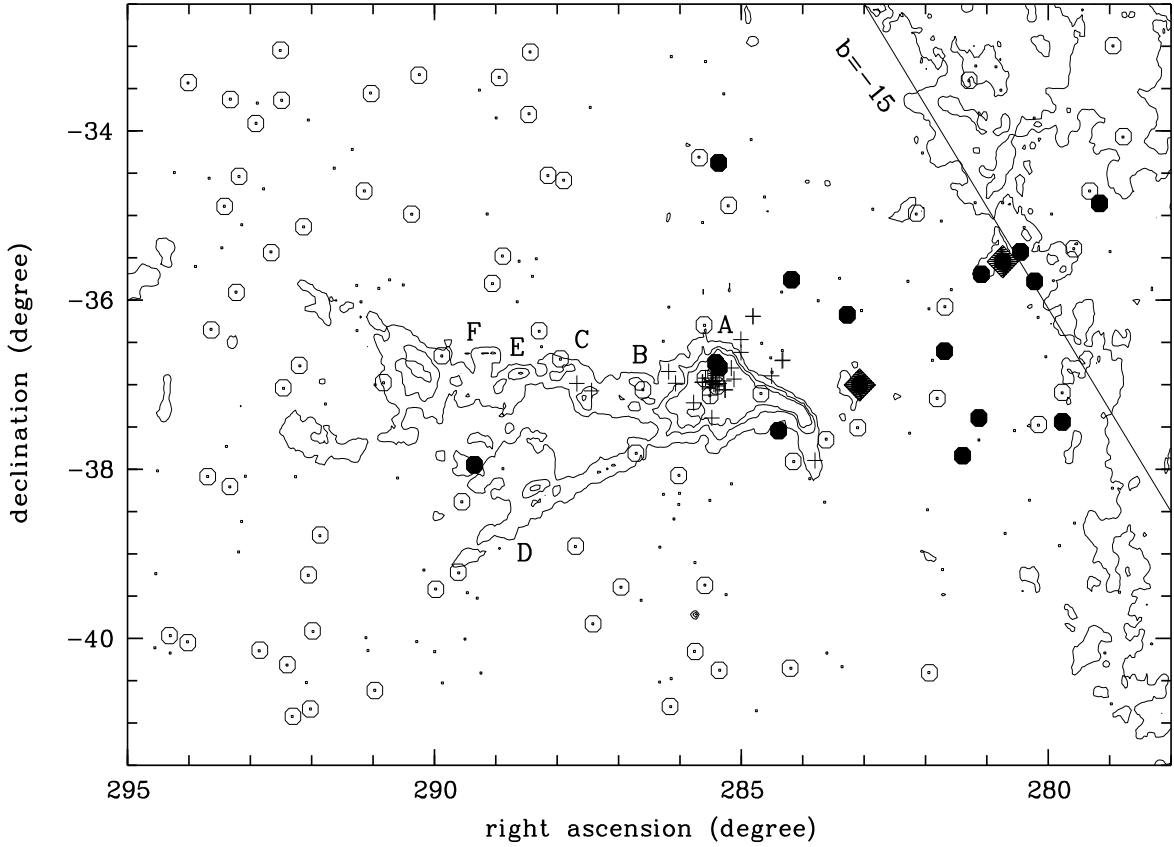,width=18cm,height=13cm,angle=270}}
\caption{ {\bf Spatial distribution of RASS sources in CrA.}
We show all 206 RASS sources as dots superimposed on 
three contour levels of the IRAS 100~$\mu m$ map.
RASS sources for which we did optical follow-up observations
are circled (Table 3). Our new wTTS are shown by filled circles, 
new cTTS by filled diamonds (Table 4). Previously known PMS stars 
are indicated by plusses (Table 1).
The clustering in the center is around the R CrA dark cloud.
We also indicate by capital letters the condensations identified by
Rossano (1978). Both new cTTS are located outside the main clouds, 
but near small cloud-lets. The galactic parallel $b=-15^{\circ}$ 
is indicated in the upper right corner}
\end{figure*}

\begin{figure*}
\caption{ {\bf Finding charts for the new PMS stars.} 
We show $5' \times 5'$ DSS charts (north is up,
east to the left) for all new PMS stars, with one exception, 
namely RXJ1917.4$-$3756, which is easy to identify as (by far) 
the brightest star in its field (ask rne@mpe.mpg.de for charts)}
\end{figure*}

\begin{figure*}
\vbox{\psfig{figure=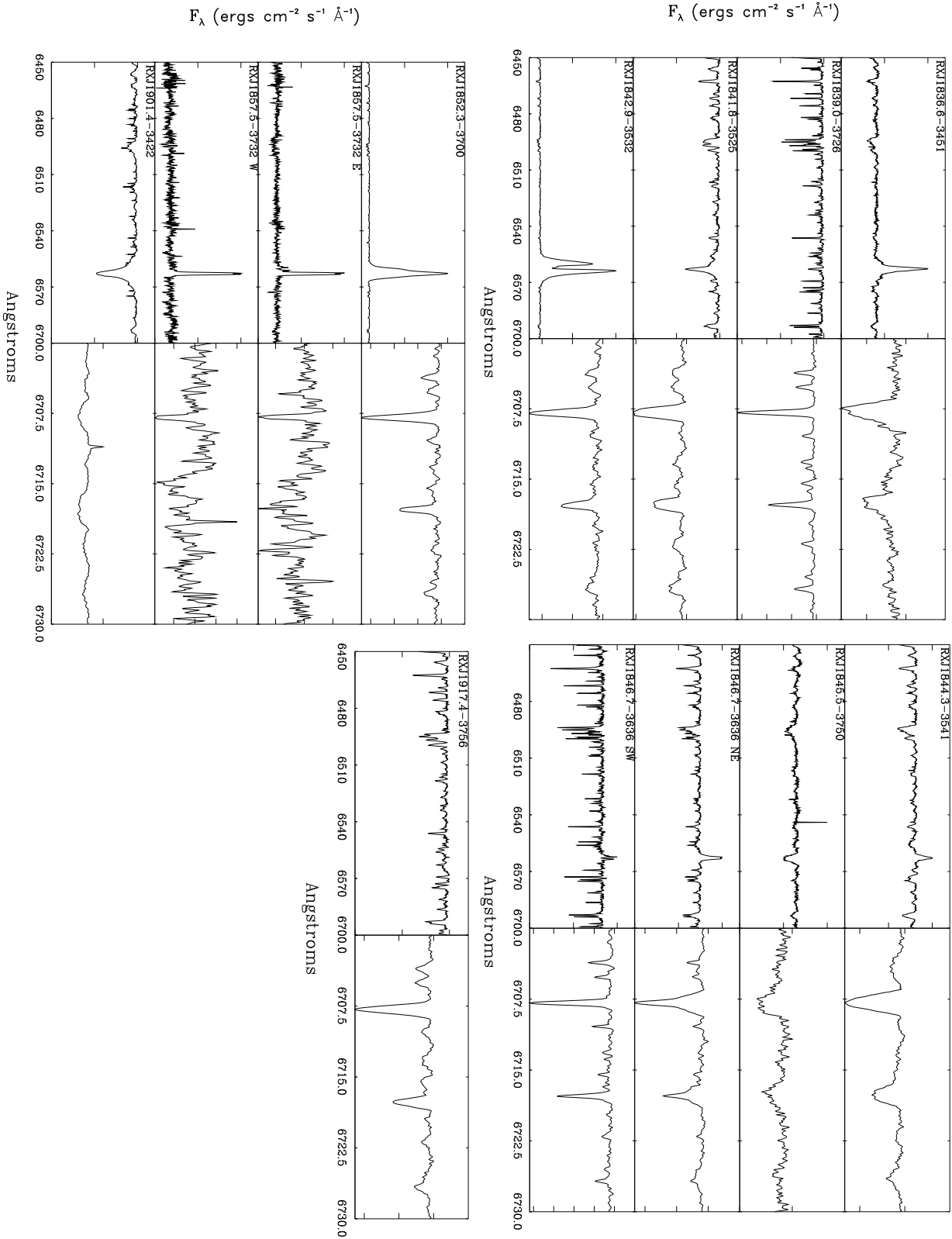}}
\caption{ {\bf High-resolution optical spectra of the new TTS.} 
The spectra of new TTS observed with the CTIO 4-m echelle. The left
panel shows a 250\AA\ region including H$\alpha$; the right panel shows
30\AA\ including the Li~I $\lambda$6707\AA\ and Ca~I $\lambda$6717\AA\
lines. The left hand panels are scaled from 0 to 120\% of the maximum 
flux, with ticks at the 50 and 100\% levels. The right hand panels are scaled 
to show the shallow absorption profiles of the rapid rotators. In the right hand
panels the ticks lie at 20\% of the continuum level, with the top of the plot
at the 120\% level. If only a single tick is shown, the bottom of the panel is
80\% of the maximum flux }
\end{figure*}

\begin{figure*}
\vbox{\psfig{figure=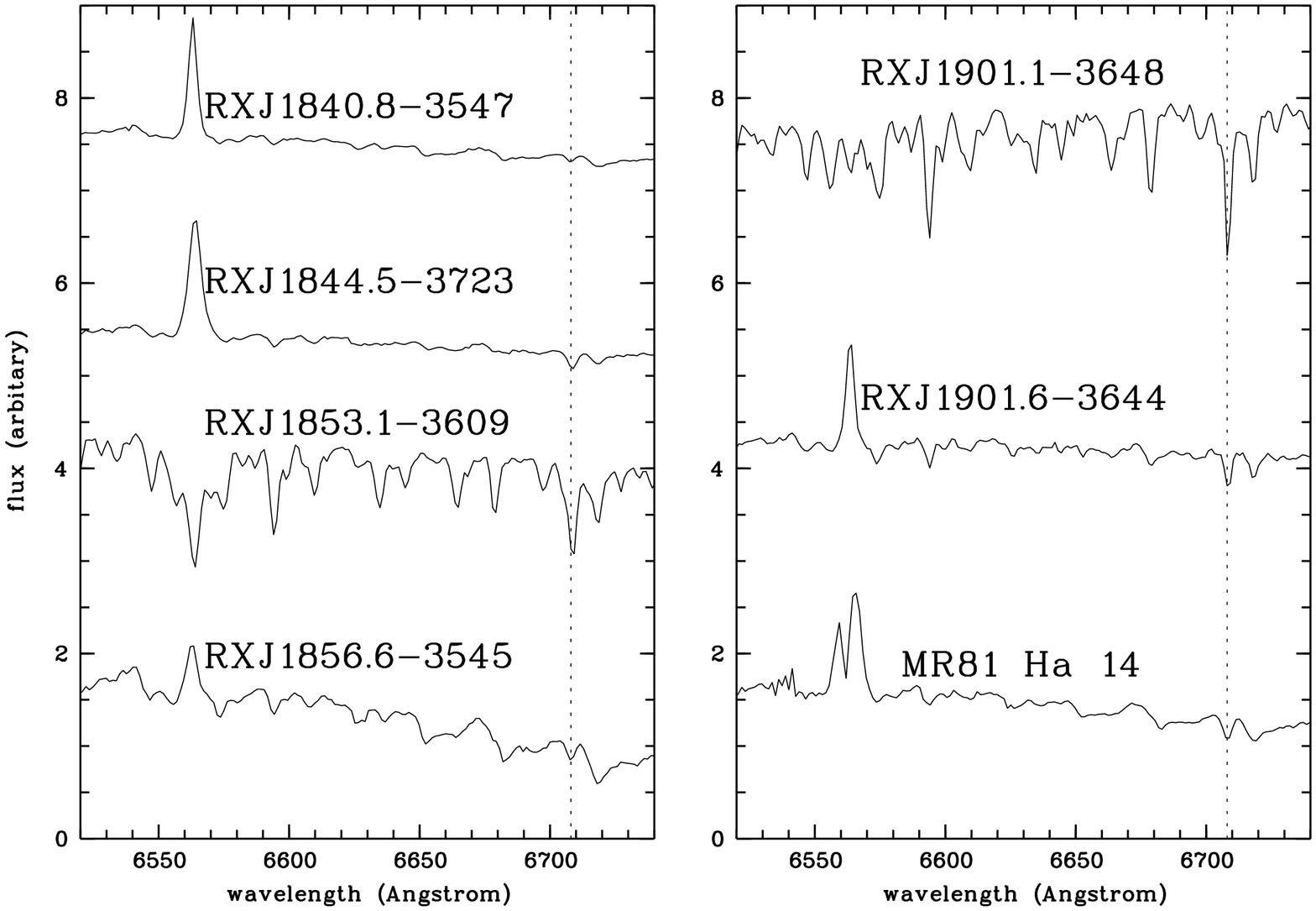,width=18cm,height=11cm,angle=270}}
\caption{ {\bf Low-resolution optical spectra of the other new TTS.} 
Spectra obtained at the ESO-1.52m of those new TTS not shown in Fig. 3
as well as for MR81 H$\alpha$ 14, also found to be wTTS. The dotted 
line shows the location of the lithium 6708 \AA~line }
\end{figure*}

In Table 3, we list the nature of the newly identified X-ray sources.
Among the 56 X-ray sources for which we performed low-resolution spectroscopy
of potential optical counterparts, we identified 19 new pre-main sequence  
stars, nine new zero-age main-sequence stars, and seven new dKe/dMe stars.
Two of the new PMS stars have $W_{\lambda}(H\alpha)~\ge~10$\AA , i.e. are classical TTS,
and 17 are weak-line TTS, four of which form two visual pairs, which are spatially 
unresolved with ROSAT, i.e. one pair corresponds to one X-ray source each.

The spatial distribution of the CrA RASS sources and TTS is shown in Fig. 1.
While all the previously known TTS in CrA including those found by EO
cluster on the main dark cloud, the new ROSAT TTS are more widely distributed.
In particular, they are almost all located west of the cloud.
The two new cTTS appear to be situated in two small cloud-lets,
where they most certainly have formed.
The new wTTS may have also formed in such small cloud-lets outside the 
current borders of the CrA cloud, and those small cloud-lets may well
have dispersed since they formed stars, similar to some seemingly
off-cloud wTTS near Cha I (Alcal\'a et al. 1995, Mizuno et al. 1998),
see below.

Finding charts for the new PMS stars can be found in Fig. 2, 
the optical spectra for our new TTS are shown in Fig. 3 
(high-resolution spectra from CTIO, if available)
and Fig. 4 (low-resolution spectra from ESO-1.52m for the other new TTS).
In Fig. 5, we also show the four spectra of previously suspected
young stars obtained at the ESO-3.5m-NTT.

As seen in Fig. 6, all new TTS show more lithium than ZAMS stars of the same 
spectral type, i.e. are younger than ZAMS; hence, they are pre-MS stars.
There are some new M-type TTS in CrA, which have depleted more than
half of their primordial lithium (lower right of Fig. 6).
As for similar M-type TTS in Taurus, these objects could
be classified as post-TTS.

\begin{figure}
\vbox{\psfig{figure=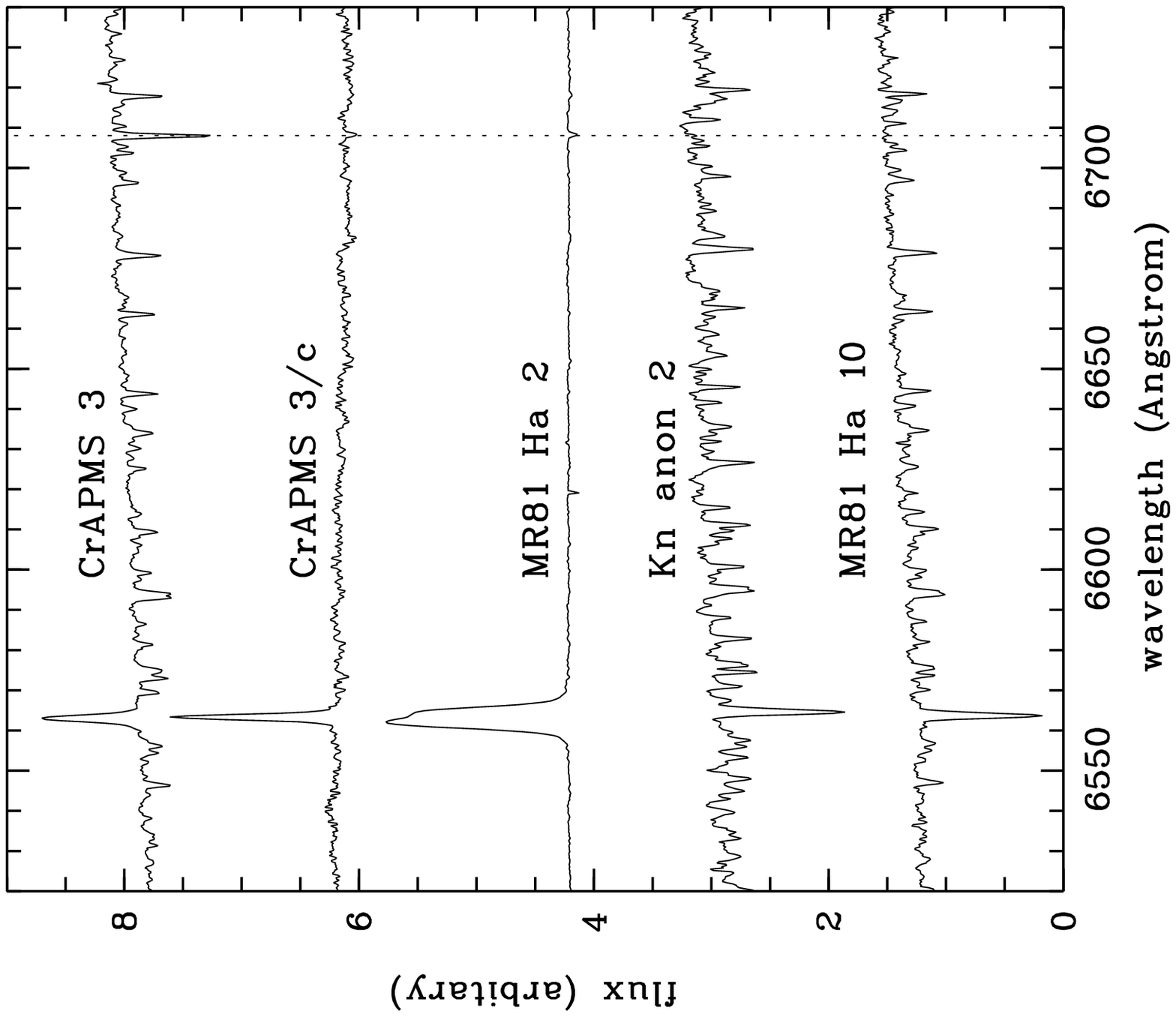,width=15cm,height=10cm,angle=270}}
\caption{ {\bf Low-resolution optical spectra of previously suspected young stars.} 
Spectra obtained at the ESO-3.5m-NTT of stars suspected to be young prior to
the ROSAT mission, namely CrAPMS 3/c (new wTTS), MR81 H$\alpha$ 2 (new cTTS),
Kn anon 2, MR81 H$\alpha$ 10 (no lithium, i.e. both non-TTS), and for comparison
we also show CrAPMS 3 (wTTS). The dotted line shows the location of
the lithium 6708 \AA~line}
\end{figure}

\begin{figure*}
\vbox{\psfig{figure=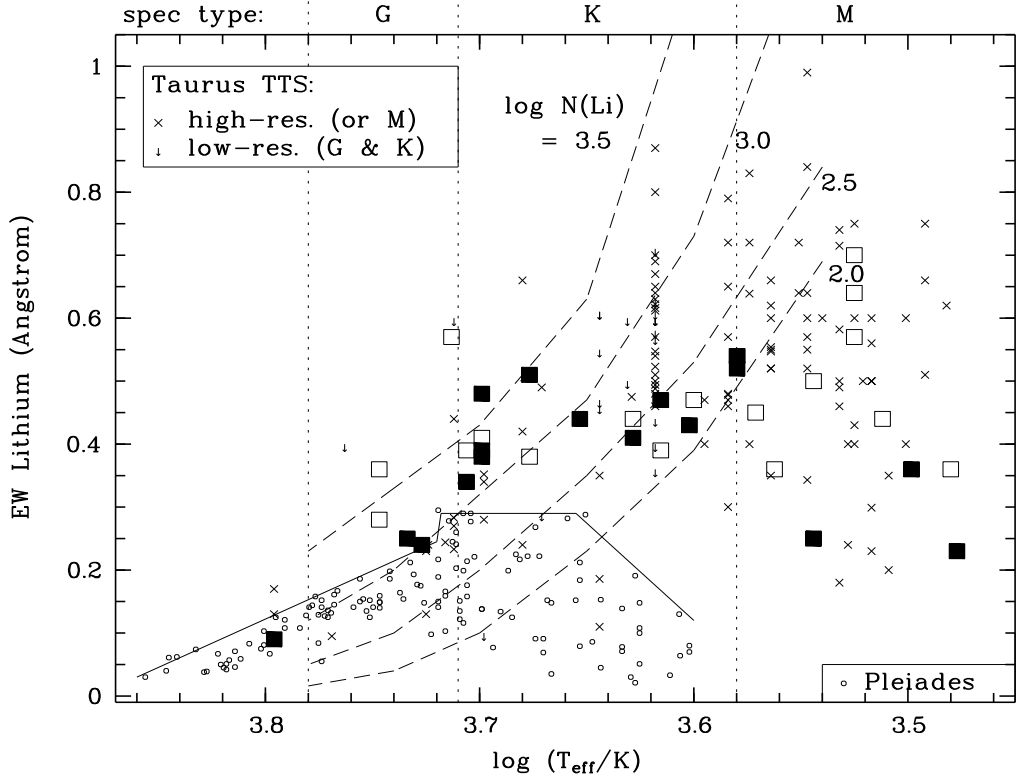,width=18cm,height=12cm,angle=270}}
\caption{ {\bf Lithium versus effective temperature.}
We plot the lithium equivalent width $W_{\lambda}$(Li) versus the
effective temperature $T_{eff}$ for 
the previously known TTS in CrA (Table 1, open squares), 
the newly identified PMS stars (Table 4, full squares),
bona-fide Taurus TTS (crosses and arrows, see Neuh\"auser et al. 1997 for references),
and the Pleiades as dots (Soderblom et al. 1993, Garc\'\i a L\'opez et al. 1994). 
We converted spectral types to $T_{eff}$ following Bessell (1979, 1991). 
Also shown are lithium iso-abundance lines (as dashed lines) for 
$\log~g~=~4.5$ from Pavlenko \& Magazz\`u (1996).
Stars with more lithium than ZAMS stars of the same spectral type, 
i.e. stars which lie above the upper envelope (solid line) of the 
Pleiades lithium data, are younger than ZAMS, i.e. PMS stars. 
Because late F- and G-type stars do not significantly burn lithium during 
the ZAMS phase, the three full squares in the lower left could be either
PMS or ZAMS stars }
\end{figure*}

\begin{table*}
\label{tab4new}
\caption []{ {\bf New T~Tauri stars in CrA.} \\
We list object number and ROSAT designation as in Tables 2 and 3,
pre-main sequence type, spectral type, H$\alpha$ and lithium equivalent 
widths (negative when in emission) from high-resolution spectra (unless given with colons), 
rotational velocity, helio-centric radial velocity, X-ray luminosity (in 130~pc 
distance, calculated from counts, exposure and ECF listed in Table 2,
65~pc for TTS no. 82), and some remarks.
}

\vspace{-.5cm}

\begin{tabular}{rlccrrrrrl} \\ \hline
No.   & Object      & PMS  & Spec & $W_{\lambda}$(H$\alpha$) & $W_{\lambda}$(Li) 
& $v \cdot \sin i$ & RadVel & $\log~L_{X}$ & Remarks \\
{\scriptsize Tab.2} & designation & type & type & [\AA ] & [\AA ] 
& $[km~s^{-1}]$ & $[km~s^{-1}]$ & $[erg~s^{-1}]$ & \\ \hline

8   & RXJ1836.6-3451 & wTTS & M0  &$ -2.4 $ & 0.54 & 34.1 &$-1.6$& 30.27 & \\
13  & RXJ1839.0-3726 & wTTS & K1  &$ +0.1 $ & 0.34 & 17.8 &$-4.8$& 30.24 & (1) \\
19  & RXJ1840.8-3547 & wTTS & M4  &$ -6.9:$ & 0.36:&      &      & 29.82 & \\
23  & RXJ1841.8-3525 & wTTS & G7  &$ +1.0 $ & 0.25 & 30.0 &$-3.1$& 30.62 & (2) \\
27  & RXJ1842.9-3532 & cTTS & K2  &$-30.7 $ & 0.38 & 23.7 &$-1.2$& 30.34 & (3) \\
30  & RXJ1844.3-3541 & wTTS & K5  &$ -0.4 $ & 0.41 & 38.8 &$-4.9$& 30.60 & \\
31  & RXJ1844.5-3723 & wTTS & M0  &$ -5.3:$ & 0.52:&      &      & 30.59 & \\
36  & RXJ1845.5-3750 & wTTS & G8  &$ +0.8 $ & 0.24 & 54.6 &$-1.3$& 30.81 & HD 173148 (1) \\
39  & RXJ1846.7-3636~NE & wTTS &K6&$ -0.6 $ & 0.47 & 30.8 &$-2.5$& 30.38 & 8" pair (4) \\
39  & RXJ1846.7-3636~SW & wTTS &K7&$ -0.2 $ & 0.43 & 17.5 &$-2.6$& 30.38 & 8" pair (4) \\
53  & RXJ1852.3-3700 & cTTS & K3  &$-33.8 $ & 0.51 & 21.8 &$ 0.0$& 30.41 & IRAS 18489-3703 (5) \\
55  & RXJ1853.1-3609 & wTTS & K2  &$ +1.4:$ & 0.39:&      &      & 30.60 & HD 174656 \\
64  & RXJ1856.6-3545 & wTTS & M2  &$ -1.6:$ & 0.25:&      &      & 29.78 & \\
68  & RXJ1857.5-3732~E&wTTS & M5  &$ -3.4 $ & 0.23 & 28.7 &$-3.1$& 29.93 & close binary (4) \\
68  & RXJ1857.5-3732~W&wTTS & M6  &$ -4.7 $ & 0.42 & 11.9 &$-2.9$& 29.93 & close binary (4) \\
80  & RXJ1901.1-3648 & wTTS & K4  &$ +0.1:$ & 0.44:&      &      & 29.88 & VSS VIII-27 \\
82  & RXJ1901.4-3422 & wTTS & F7  &$ +2.2 $ & 0.09 & 57.6 &$-3.3$& 29.77 & HD 176383 (2) \\
85  & RXJ1901.6-3644 & wTTS & M0  &$ -2.3:$ & 0.54:&      &      & 30.18 & VSS VIII-26 \\
135 & RXJ1917.4-3756 & wTTS & K2  &$ -0.9:$ & 0.48:&      &      & 30.61 & SAO 211129 \\ \hline

\end{tabular}

Notes:
(1) Complex H$\alpha$ profile.
(2) Given the relatively early spectral type and relatively small lithium 
strength, this star could be either pre- or zero-age MS stars (see also Sect. 6.3).
Its Hipparcos distance is 65 pc, $L_{X}$ given for 65 pc.
(3) Also observed with low resolution in 1995 and twice in 1996, when
$W_{\lambda}(H\alpha)$ varied from $-53.6$\AA~to $-88.7$\AA~from night to night.
Double-peaked $H \alpha$ emission, also HeI and [OII] emission.
(4) Visual pair; position given in Table 3 is for the primary, i.e. the slightly
brighter star; not resolved spatially with ROSAT (the combined $L_{X}$ is given).
(5) Observed once in 1995 with low resolution and about the same
H$\alpha$ as given above for high-resolution spectrum,
and five times with low resolution in 1996 with 
$W_{\lambda}(H\alpha)$ from $-5.22$\AA~to $-8.97$\AA .

\end{table*}

\section {Optical and infrared photometry}
\label{yso}

\begin{table*}
\label{tab_ph}
\caption []{ {\bf Photometric data for new T~Tauri stars in CrA.} \\
We list object number as in Tables 2 to 4.
Values in brackets are errors for the last digit(s).
When no error is given, data are taken from Simbad, GSC, or USNO,
which are more uncertain (indicated by colons).
For TTS no. 64, the JHK data are from the 2nd 2MASS data release.
Bolometric luminosity are estimated for a distance of 130 pc 
(65 pc for TTS no. 82). Ages and masses are estimated by comparison 
with isochrones and tracks from D'Antona \& Mazzitelli (1994).
The last object, RXJ1855.1-3754, is the TTS found in Neuh\"auser et al. 
(1997), a ROSAT HRI source, not a RASS source, see Table 1. }

\vspace{-.5cm}

\begin{tabular}{lccccccccrcc} \\ \hline
No.    & B   &  V & V--R & R--I & K & J--K & H--K & $A_{V}$ & $\log$ & age & mass \\
Tab. 2 & mag & mag & mag & mag & mag & mag & mag & mag & $L_{bol}/L_{\odot}$ & $[Myr]$ & $[M_{\odot}]$ \\ \hline

8   & 14.3: & 13.22 (3) & 0.98 (1) & 1.08 (1) & 9.02 (5) & 0.86 (6) & 0.20 (6) & 0.7 & $-0.45$ & 3.2 & 0.6 \\
13  & 12.0: & 10.81 (1) & 0.50 (2) & 0.50 (1) & 8.51 (2) & 0.62 (3) & 0.14 (3) & 0.2 & 0.04 & 10 & 1.3 \\
19  & 18.8: & 14.99 (3) & 1.24 (3) & 1.34 (2) &10.14 (4) & 0.91 (16)& 0.25 (7) & 0.0 & $-0.68$ & 1.6 & 0.3 \\
23  &      & 9.7:  & &      & 8.04 (4) & 0.41 (15)& 0.07 (7) & 0.0 & 0.30 & 10 & 1.4 \\
27  & 12.3: & 12.28 (3) & 0.70 (1) & 0.71 (1) & 8.23 (3) & 1.35 (5) & 0.51 (5) & 1.5 & $-0.04$ & 10.0 & 1.2 \\
30  &      & 11.3:    & &      & 8.38 (3) & 0.71 (5) & 0.18 (5) & 0.1 & 0.07 & 1.0 & 0.9 \\
31  & 14.3: & 13.1:   & &      & 9.38 (3) & 0.85 (6) & 0.22 (5) & 0.1 & $-0.64$ & 6.3 & 0.6 \\
36  & 10.0: & 9.2:  & &      & 7.18 (3) & 0.44 (5) & 0.10 (5) & 0.3 & 0.73 & 4.0 & 1.5 \\
39-NE &   & 11.16 (3) & 0.73 (2) & 0.66 (1) & 7.89 (3) & 0.72 (5) & 0.11 (5) & 0.4 & 0.25 & 0.5 & 0.7 \\
39-SW &   & 11.43 (2) & 0.24 (1) & 0.66 (1) & 7.89 (3) & 0.72 (5) & 0.11 (5) & 0.5 & 0.20 & 0.5 & 0.6 \\
53  & 12.7: & 12.35 (3) & 0.73 (1) & 0.72 (1) & 9.01 (3) & 0.78 (5) & 0.21 (5) & 1.0 & $-0.25$ & 10 & 1.1 \\
55  & 11.0: & 9.6:  & &      & 7.31 (3) & 0.61 (5) & 0.12 (5) & 0.1 & 0.46 & 2.0 & 1.7 \\
64  & 14.9: & 12.9: & & & 9.91 (3) & 0.91 (5) & 0.30 (7) & 0.0 & $-0.25$ & 0.8 & 0.4 \\
68-E& & 15.56 (3) & 1.2 (1) & 1.5 (1) & 10.53 (2) & 0.96 (4) & 0.20 (3) & 0.0 & $-0.91$ & 2.5 & 0.2 \\
68-W& & 16.5 (3) & 1.2 (3) & 1.7 (2) & 11.28 (3) & 0.95 (5) & 0.31 (4) & 0.0 & $-1.29$ & 2.5 & 0.1 \\
80  &   & 15.85 (4) & 1.29 (3) & 1.52 (2) & & & & 3.0 & $-0.86$ & 40 & 0.7 \\
82  & 8.7:  & 8.2:  & &      & 7.07 (3) & 0.23 (5) & 0.05 (5) & 0.0 & 0.30 & 30 & 1.2 \\
85  & 15.5: & 14.27 (3) & 1.17 (2) & 1.23 (2) & & & & 1.1 & $-0.70$ & 7.9 & 0.6 \\
135 & 10.8: & 9.9:  & &      & 7.54 (3) & 0.69 (5) & 0.14 (5) & 0.1 & 0.36 & 3.2 & 1.6 \\ \hline
\multicolumn{2}{l}{RXJ1855.1-3754} & 13.05 (5) & 0.62 (3) & 0.57 (3) & & & & 0.3 & $-0.82$ & 40 & 0.7 \\ \hline

\end{tabular}

\end{table*}

Broad band photometric observations in the VRI filters were carried out in 
1997 from June 22 to 30 using the 0.9m telescope of Cerro Tololo Interamerican 
Observatory (CTIO). The detector was a CCD Tektronix 2048 (CTIO \# 3) with pixel 
size 24 $\mu$m and a readout noise of 3 to 5 electrons. 
The whole CCD was read out. Dome flat-field exposure sequences in each 
filter were taken typically before the beginning and upon the end of every 
night for flat field correction. Two or three different standard star fields 
from Landolt (1992) were also observed every night at different 
airmasses for the determination of atmospheric extinction, zero points and 
color transformation to the Johnson-Kron-Cousins standard system.

Raw CCD frames were bias subtracted and flat fielded using the IRAF
package CCDRED. Sky flats were used, and the flat field variation 
across the final images was spot-checked and found to be negligible.  
The photometric solution for each night was determined using the IRAF
tasks APPHOT and PHOTCAL. 
First, an aperture for each night was chosen by inspecting the
reduced images from each night. The average FWHM over the run was
about 1.3$^{\prime\prime}$. To insure that all the light from a given
star was in the aperture, a typical aperture for source extraction 
was taken as four times the FWHM or about 6$^{\prime\prime}$.
Annuli for background subtraction varied but were usually about
2$^{\prime\prime}$ from inner diameter to outer diameter with an 
inner diameter of about 8$^{\prime\prime}$. The background was taken
as the median value within the annuli. An average of 30
calibration standards were observed each night at various airmasses.
Photometric solutions for each night were determined by fitting the
data to a color dependent airmass equation.
The photometric errors dominate the Poisson errors in most cases.

Once the photometric solution for each night was determined, the target
fields were examined.  Aperture photometry was performed on all stars
using the same aperture as was used for the standard stars on that
night. The only exception to this was RXJ1857.5-3732.  
In this case, the two stars were separated by less 
than 4$^{\prime\prime}$.  Therefore we
used the IRAF tool SUBSTAR to subtract one star from the image so we
could accurately measure the instrumental magnitude of the other. 
The observed magnitude and colors were obtained by applying the
instrumental magnitudes derived from the aperture to the photometric
equation for the given night.
Results are listed in Table 5.

The near-IR data were obtained using the CTIO Infra-Red Imager,
CIRIM\footnote{see Elston 1999 at http://www.ctio.noao.edu/ 
instruments/ir$_{-}$instruments/cirim/cirim.html}, 
on the 1.5m telescope at CTIO. CIRIM is a 256 $\times$ 256
HgCdTe array. We observed at the f/13.5 focus, giving a plate scale of
0.65 arcsec per pixel. The data were obtained on the nights of 1998 July
8 to 14. All nights were photometric. The stellar
point-spread function is well-sampled photometrically.
Dome flats were taken at the start of each night.
Flats were obtained with the dome illumination lamps both on and off,
to determine the thermal contribution to the flat image.
Standard stars, taken from the lists of Elias (1982) and the UKIRT faint
standards (Casali \& Hawarden 1992),
were observed hourly, at a full range of airmasses. Exposure times
varied between 0.4 and 3 seconds, based on source magnitude,
with 3 images coadded at each position. We observed the standard stars
using a 2 $\times$ 2 raster, with 30 arcsec spatial offsets between positions.
We observed the targets using exposure times between 0.4 and 20 seconds,
based on expected source
brightness, with 3 images coadded at each position. We observed
using a 2 $\times$ 3 raster, with spatial offsets of 15 arcsec between frames.

We linearized the data using the IRAF routine IRLINCOR. All other processing
was undertaken using our IDL-based CIRIM reduction 
package\footnote{http://sbast3.ess.sunysb.edu/fwalter/CIRIM/cirim.html}.
The images are divided by the appropriate normalized flat field image.
The images in the raster pattern are median-filtered to determine the
local sky image, which is then subtracted from each flattened image. The
images in the raster pattern are then aligned by cross-correlating on the
brightest sources in the image, and co-added. The image center and plate
scale are determined
by cross-correlating the images with stars in the USNO catalog (Monet
et al. 1998). The photometric solution is determined by fitting the log of
the standard star counts within a 12-pixel (7.8 arcsec) radius region
as a linear function of the air mass.
The solution is edited interactively to remove discrepant stars and/or points.
The RMS scatter in the photometric solutions is $1$ to $2\%$, and we take 
this to be our photometric precision.
Magnitudes are determined by applying the photometric solution to the
net counts observed in a 12-pixel (7.8 arcsec) radius region. In the case of
close binaries, we determined the total flux using a large extraction radius,
and the relative magnitudes using smaller extraction radii. Typical
uncertainties are $\pm 0.02$~mag, but this degrades for the fainter targets.
Results are listed in Table 5.

\section{H-R diagram, ages, and masses}

Because we have a homogeneous and almost complete set of precise VJHK photometric
data, we estimate the absorption in the line-of-sight to the stars from the observed 
$V-J$, $V-H$, and $V-K$ colors and their intrinsic color indexes, which we know from 
their spectral types. For those few stars, for which JHK is not available, we 
estimate the absorption from VRI colors. Visual extinctions $A_{V}$ for all new TTS 
are listed in Table 5. Then, we can also estimate the bolometric luminosity
$L_{bol}$ at an assumed distance of 130 pc, also listed in Table 5.

\begin{figure}
\label{hrd}
\vbox{\psfig{figure=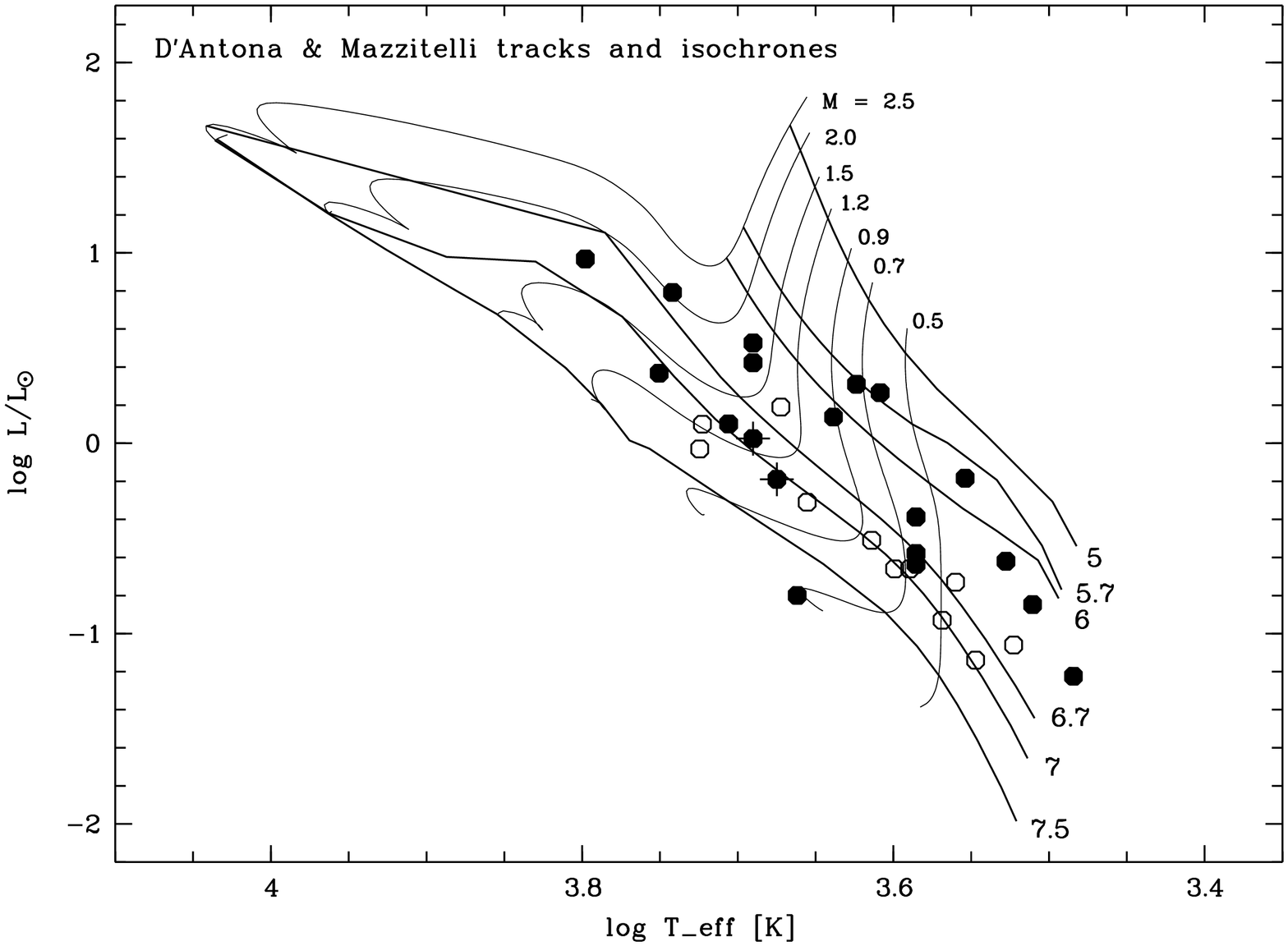,width=9cm,height=7cm,angle=270}}
\caption{ {\bf H-R diagram with the newly found young stars.} 
Bolometric luminosity versus effective temperature for all newly
found Li-rich ROSAT counterparts (from Table 4, all filled symbols)
and for Li-rich EO counterparts (Walter et al. 1997, open symbols),
compared to tracks and isochrones from D'Antona \& Mazzitelli (1994)
for ages (given on a log scale) and masses (given in $M_{\odot}$).
The two newly found cTTS are marked by a plus sign;
surprisingly, they are relatively old, namely $\sim 10$ Myrs old
(at the assumed distance of 130 pc)}
\end{figure}

We display the location of the new TTS in the H-R diagram in Fig.~7
together with tracks and isochrones from D'Antona \& Mazzitelli (1994).
They yield rough estimates for ages and masses of our new TTS,
which we also list in Table 5.
It can been seen in Fig.~7 that the two visual pairs
of TTS, RXJ1846 and RXJ1857, are both roughly co-eval.
The ages of the new TTS range from less than one million years to 
a few tens of million years, not surprising given the wide
spatial distribution. From that wide distribution on the sky around
the CrA dark cloud, we have to conclude that the distances can show
a similar spread, so that the ages and masses are uncertain.
For the newly found young star RXJ1901.4-3422, whose parallax was
measured by the Hipparcos satellite, we estimated its luminosity
using the Hipparcos distance of 65 pc; it then falls on the
30 Myrs isochrone, just above the ZAMS; because its spectral type 
is late F, it is difficult to
classify it as either pre- or zero-age MS given its lithium.
One more newly found Li-rich star, RXJ1901.1-3648, is also found
near the ZAMS, using 130 pc as distance, which may of course be a
wrong distance. The latter one has spectral type K4 and clearly more
lithium than Pleiades K4-type stars; it is located right on the cloud
and its extinction is large ($A_{V} \simeq 3$ mag, Table 5), hence it may be 
deep inside the dark cloud. 

Our new Li-rich ROSAT counterparts show a larger age spread than
the Li-rich EO counterparts (Walter et al. 1997). This is probably 
due to the larger spread in distances among the ROSAT stars compared 
to the EO stars, because we investigated a larger area on the sky 
using the RASS as compared Walter et al. (1997) who used EO pointed 
observations centered on the CrA dark cloud.

The two new cTTS, RXJ1842.9-3532 and RXJ1852.3-3700, both
appear to be $\sim 10$ Myrs old, which is relatively old for cTTS.
This result only holds if they are indeed at $\sim 130$ pc as
assumed for calculating their $L_{bol}$, which may be unrealistic,
because they are located on the line-of-sight to two small cloud-lets,
both off the main CrA dark cloud. On the other hand, there are also
some other cTTS with ages around 10 Myrs, namely TW Hya and HD 98800,
for which the Hipparcos parallaxes are known, so that they could be
placed correctly into the H-R diagram. 
Both two new CrA cTTS show only small IR excess emission at JHK,
and one of them, RXJ1852.3-3700, is an IRAS source.

The two TTS with the latest spectral types in our sample are 
RXJ1857.5-3732 E and W with M5 and M6, respectively. They are both 
certainly young because of the strong lithium absorption.
Hence, they are most certainly members of the CrA dark cloud,
i.e. at 130 pc. Then, in the H-R diagram, they lie near the borderline 
between stars and brown dwarfs with masses around $\sim 0.1$ 
to $\sim 0.2~M_{\odot}$. At first glance, it appears surprising 
to find such low-mass objects among RASS sources. 
Previously, several borderline objects and even brown dwarfs were 
detected as X-ray sources, but only in very deep pointed observations
(Neuh\"auser \& Comer\'on 1998, Neuh\"auser et al. 1999,
Neuh\"auser \& Comer\'on 2000).
However, our two RASS detected mid- to late-M dwarfs form
a close pair, i.e. only one unresolved RASS X-ray source.
Given their optical/IR luminosities, spectral types, and typical
$L_{X}/L_{bol}$ ratios, they would not have been detected 
individually in the RASS.

\section{Kinematics}
\label{pmr}

Because CrA is not part of the Gould Belt, the young stars surrounding 
the CrA dark cloud most certainly are associated with the CrA association. 
They could have formed near the present locations, where the gas has 
dispersed since then. In that case, these young stars should be somewhat 
older than the on-cloud TTS. Alternatively, if the CrA cloud has not shrunk,
the off-cloud young stars could have dispersed out of the cloud,
either slowly (then, the outermost stars would again be older than the
on-cloud stars) or, at least in some cases, have higher velocities. 
In the latter case, one would expect at least a 
few young run-away TTS ejected from the dark cloud.
For checking these possibilities, we need to
investigate proper motions and radial velocities.

\subsection{Proper motions}

In order to examine the kinematical state of the stars in the R~CrA association,
we searched for proper motions for all the stars in Tables 1 and 3 in the
Hipparcos (ESA 1997), PPM (R\"oser \& Bastian 1991, Bastian et al. 1993,
R\"oser et al. 1994), ACT (Urban et al. 1997), TRC (H{\o}g et al. 1998),
Tycho2 (H{\o}g et al. 2000),
and STARNET (R\"oser 1996) proper motion catalogs. Altogether we could identify
24 stars in these catalogs, including nine 
stars known already before the ROSAT
mission, nine TTS newly identified here, as well as four ZAMS and two dKe/dMe stars.
Note that R~CrA itself is not listed; it is included in the
Hipparcos catalog (HIP~93449), but no meaningful solution for its astrometric
parameters could be derived. Proper motions are given in Table~\ref{pm_table}. 
For stars present in more than
one catalog we usually adopted the most precise proper
motion determination, unless it was in conflict with values from other
catalogs. All proper motions were transformed to the Hipparcos astrometric
system (ICRS) before comparison.

\begin{table}
\begin{center}
\caption[]{\label{pm_table} {\bf Proper motions for CrA TTS.}\\
Listed are objects from Tables 1 to 4, which are included
in at least one of the proper motion catalogues Hipparcos (HIP), PPM,
ACT, TRC (T), STARNET (S) or Tycho2 (T2). All proper motions refer to the 
Hipparcos astrometric system.}
\begin{tabular}{@{}llr@{$\pm$}lr@{$\pm$}l@{}}
\hline\noalign{\smallskip}
 & &  
\multicolumn{2}{c}{$\mu_{\alpha}\cdot\cos\delta$} & 
\multicolumn{2}{c}{$\mu_{\delta}$} \\
\raisebox{1.5ex}[-1.5ex]{object} & 
\raisebox{1.5ex}[-1.5ex]{cat.\ no.} &
\multicolumn{2}{c}{[mas$\cdot$yr$^{-1}$]} & 
\multicolumn{2}{c}{[mas$\cdot$yr$^{-1}$]} \\
\noalign{\smallskip}\hline\noalign{\smallskip}
\multicolumn{6}{l}{Known and suspected CrA member stars before ROSAT:} \\
\noalign{\smallskip}\hline\noalign{\smallskip}
185720-3643       & T2 7421 1242 &   5.3 & 2.6 & --29.1 & 2.7 \\
185801-3655       & S  7421 1040 &  13   & 6.1 & --27   & 6.1 \\
HR 7169$^{a}$     & T  7421 2294 &  10.1 & 2.6 & --28.8 & 2.6 \\
HR 7170$^{a}$     & T  7421 2295 &   7.0 & 2.6 & --15.5 & 2.6 \\
S CrA             & S  7421  213 &  10   & 4.6 & --26   & 4.6 \\
CoD--37\degr13022 & T2 7421 1890 &   6.6 & 2.5 & --27.6 & 2.6 \\
HD 176386         & HIP 93425    &   1.6 & 1.8 & --26.7 & 0.9 \\
V702 CrA          & T2 7421  493 &   4.9 & 1.8 & --26.1 & 1.8 \\
SAO 210888        & HIP 93689    &   4.5 & 1.6 & --28.6 & 0.8 \\
\noalign{\smallskip}\hline\noalign{\smallskip}
\multicolumn{6}{l}{T Tauri stars newly discovered with ROSAT (RXJ...):} \\
\noalign{\smallskip}\hline\noalign{\smallskip}
1839.0-3726       & T2 7419   76 &   0.6 & 2.6 & --29.1 & 2.6 \\
1841.8-3525       & T2 7415  284 &   7.0 & 1.9 & --25.1 & 1.9 \\
1842.9-3532       & S  7415  696 &   3   & 4.8 & --27   & 4.8 \\
1844.3-3541       & S  7419   22 & --8   & 4.8 & --32   & 4.8 \\
1845.5-3750       & T2 7915  531 &   4.6 & 2.0 & --25.9 & 2.0 \\
1852.3-3700       & S  7420  939 &   1   & 5.8 & --28   & 5.8 \\
1853.1-3609       & T2 7420  774 &   2.9 & 1.8 & --24.6 & 1.7 \\
1901.4-3422$^{b}$ & HIP 93412    &   9.9 & 1.5 & --46.8 & 1.0 \\
1917.4-3756       & T2 7918  222 &   8.4 & 1.5 & --25.8 & 1.5 \\
\noalign{\smallskip}\hline\noalign{\smallskip}
\multicolumn{6}{l}{Other stars (RXJ...):} \\
\noalign{\smallskip}\hline\noalign{\smallskip}
1840.6-3728       & T2 7419   58 & --2.5 & 2.8 &  --6.4 & 2.9 \\
1911.5-3434       & T2 7418 2446 &  13.3 & 1.8 & --15.7 & 1.9 \\
1915.7-3321       & S  7427 1333 & --3   & 5.1 &  --5   & 5.1 \\
1921.4-3459$^{c}$ & HIP 95149    &  78.9 & 4.1 &--108.9 & 2.5 \\
1928.5-3508       & HIP 95753    &--16.3 & 2.1 & --14.4 & 1.4 \\
1936.0-4002       & T2 7936  832 & --4.9 & 2.4 &    7.8 & 2.4 \\
\noalign{\smallskip}\hline\noalign{\smallskip}
\end{tabular}
\end{center}
\begin{footnotesize}
(a) The visible double star HR~7169/HR~7170 is represented by two entries in the Hipparcos
Catalogue (HIP~93368 and HIP~93371) with two individual component solutions.
The solution quality is classified as fair; however the proper motion errors 
are much smaller in TRC, which we adopt here.\\
(b) Hipparcos distance is 65$\pm$5\,pc, i.e.\ foreground to the cloud.\\
(c) RXJ\,1921.4-3459 has an acceleration solution in the Hipparcos Catalogue.
\end{footnotesize}
\end{table}

\begin{figure*}
\vbox{\psfig{figure=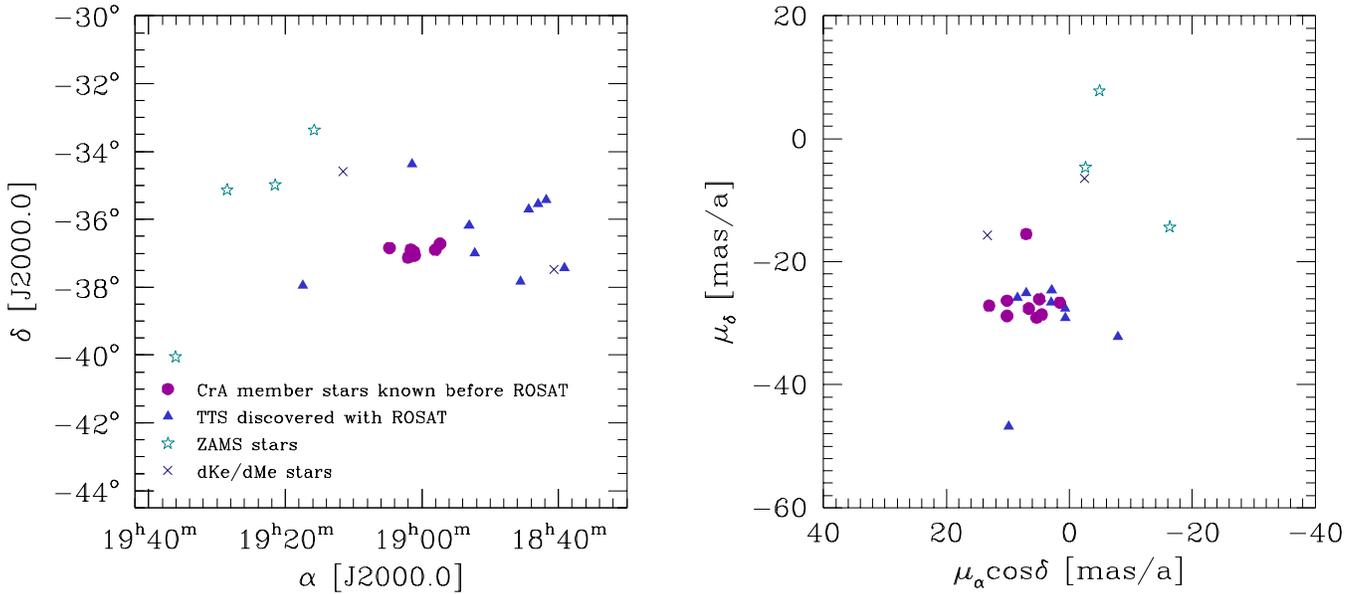,width=18cm,height=8cm}}
\caption[]{\label{pospm} {\bf Positions (left panel) and proper motions (right panel) of all
the stars in Table~\ref{pm_table}}, i.e. with proper motions available. The mean proper motion
of R~CrA member stars is well defined by most of the previously known stars (including
early-type as well as late-type stars) and newly identified young stars in this region.
Note that the proper motions of the stars classified as TTS are much
more uniformly distributed than the proper motions of the stars classified as
ZAMS or dKe/dMe  }
\end{figure*}

Positions and proper motion diagrams for stars in Table~\ref{pm_table} are
shown in Fig.~\ref{pospm}. The mean proper motion of the R~CrA member stars seems
to be very well defined. All except maybe one (HR 7170) of the member stars 
known before ROSAT (including the late B-type stars listed at the end of
Table 1, whose membership was not clear before) show very similar proper motions, 
and all except two of the newly identified TTS nicely follow this trend.
In contrast to this, the stars classified as ZAMS or dKe/dMe form a
kinematically much more inhomogeneous distribution.
The mean proper motion for all 
15 likely members of the association is
($\mu_{\alpha}\cos \delta,\mu_{\delta}$) = ($5.5,-27.0$)\,mas\,yr$^{-1}$.
The largest part of this motion is simply the reflex of the solar motion, which is
($\mu_{\alpha}\cos \delta,\mu_{\delta}$) $\approx$ ($7.1,-23.7$)\,mas\,yr$^{-1}$
at 19\hour~00\min, -37\degr~00\arcmin\ and a distance of 130\,pc, and
($\mu_{\alpha}\cos \delta,\mu_{\delta}$) $\approx$ ($5.6,-24.0$)\,mas\,yr$^{-1}$
for 18\hour~45\min and the same declination and distance values as above.
Thus the slight difference between the mean proper motion in right ascension of the
eight R~CrA member stars known before ROSAT
[\,($\mu_{\alpha}\cos \delta,\mu_{\delta}$) = ($7.0,-27.5$)\,mas\,yr$^{-1}$\,]
and the seven newly identified member TTS
[\,($\mu_{\alpha}\cos \delta,\mu_{\delta}$) = ($3.9,-26.4$)\,mas\,yr$^{-1}$\,]
is partly a projection effect reflecting the fact that the new TTS are located at
slightly lower right ascension.

\subsection{Space velocities}

\begin{figure*}
\vbox{\psfig{figure=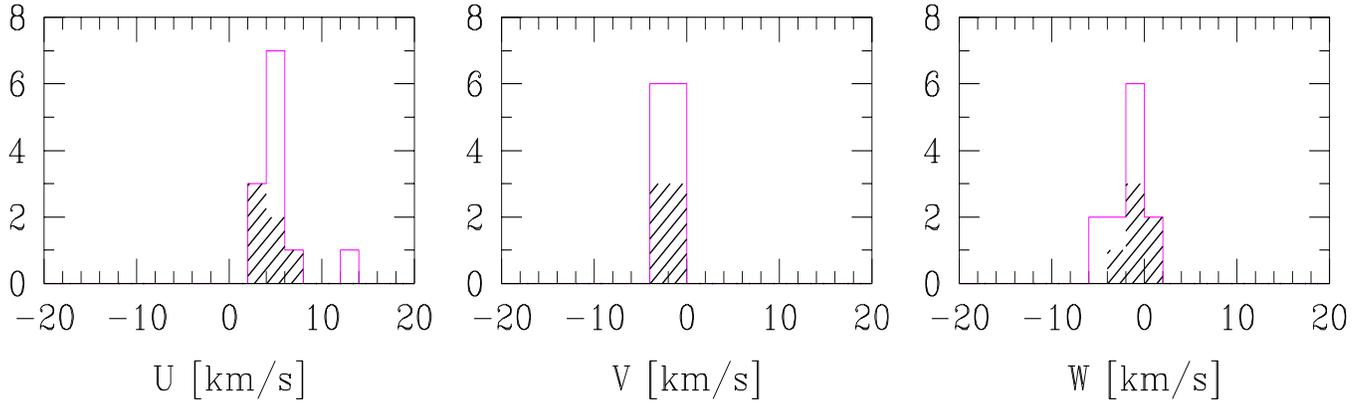,width=18cm,height=5.5cm}}
\caption[]{\label{uvw} {\bf Space velocities for all likely R~CrA member stars}
(six stars known before ROSAT and six newly identified TTS shown hatched) 
with known proper motions and radial velocities.
The effects of differential galactic rotation and the reflex of the solar
motion were eliminated using IAU standard values for all constants.
A distance of 130~pc has been assumed for all stars. U points in the
direction of the galactic center, V in the direction of galactic rotation and
W towards the north galactic pole. The velocity dispersions are very small  }
\end{figure*}

We calculated space velocities for those stars with measured radial velocities 
(taken from Table 1 for six stars and from Table 4 for another six stars) 
and corrected them for the influence of galactic rotation (Fig.~\ref{uvw}). 
The solar motion has also been subtracted, although this does
not change the relative space velocities between stars, in contrast to galactic
rotation or projection effects.

With the exception of HD~176386, which shows a discordant motion in the
U-direction, all calculated space velocities are very similar. 
The mean (U,V,W) values are ($4.8,-2.1,-2.7$)\,km\,s$^{-1}$
for the stars known before ROSAT (excluding HD~176386 for taking the mean 
of the U-ve\-lo\-ci\-ties) and ($4.4,-2.2,-0.7$)\,km\,s$^{-1}$
for the new TTS, i.e. no systematic differences seem to be
present. There could still be a small difference in the W-velocities,
but the number of stars is so small that this is maybe not significant.

The velocity dispersion is indeed very low,
($\sigma_U,\sigma_V,\sigma_W$)\,= ($1.4,1.2,1.8$)\,km\,s$^{-1}$ 
for the whole sample, which excludes the ejection mechanism 
(so-called run-away TTS) as major source for off-cloud TTS,
because they should have discrepant velocities.
The velocity dispersion is highest in the W-direction, and taking 
also into account that the positions of the stars form a 
broader distribution in the Z-direction ($\sigma_Z=4.9$\,pc, 
or $\sigma_Z=4.0$~pc if the far off lying RXJ1917.4-3756
is excluded) than in the Y-direction ($\sigma_Y=1.9$\,pc)\footnote{This can also be
seen in Fig.~1, where the Z-direction is more or less perpendicular to the
$b=-15$\degr\ line, whereas the Y-direction is parallel to this line (the new
TTS lying north of the central cloud is RXJ1901.4-3422, a young foreground
star, see Sect.~\ref{discordant}).}
this could be interpreted in terms of stars oscillating around the galactic plane.
It simultaneously would explain why no PMS stars were found in the opposite
direction of the R~CrA cloud: the stars already reached the largest
distance from the galactic plane and are currently near their turning point,
consistent with their W-velocities being close to zero. 
One complete oszillation around the galactic plane would last 
$\sim 10^{8}$ yrs, but need not be finished by now for this scenario to be true.
The off-cloud stars are currently located on average at $z = -35$~pc, 
while the cloud is at $z = -39$~pc. The stars which now appear to be off-cloud 
and the cloud itself, given their current locations and velocities and 
tracing their paths back in time, would have been at the same location
$\sim 3 \cdot 10^{7}$ yrs ago. The stars located outside of the current 
cloud borders are on average $\sim 10$ Myrs old, i.e. older than those 
inside the cloud, which supports the cloud oszillation scenario. However, 
the details and exact time-scales depend on the unknown total cloud mass.
A very similar interpretation could also explain the positions 
and the motions of the stars found south of the Taurus clouds 
(Neuh\"auser et al. 1997, Frink et al. 1997).

L\'epine \& Duvert (1994) suggested that high velocity cloud impacts could
trigger star formation, and that subsequently to the impact clouds and stars
could be separated from each other due to different friction during the passage
through the galactic plane. If this is true, the new TTS found outside the dark
cloud today could very well have been born inside the molecular cloud.
The fact that the observed velocity dispersions are so low supports this
scenario. A typical proper motion uncertainty of 3\,mas\,yr$^{-1}$ and
an assumed distance uncertainty of 15\,pc translate into a combined uncertainty
of $\approx$\,3\,km\,s$^{-1}$ for the V and W components of the space
velocity. The observed velocity dispersions are found to be even lower, so
that this is consistent with a very small intrinsic dispersion of the velocities.

\subsection{Stars with discordant proper motions}
\label{discordant}

There are three stars among the sample of pre-ROSAT R~CrA members and newly
identified TTS with discordant proper motions from the mean.

HR~7170 belongs to a possibly quadruple system, and therefore the proper motion
determination is highly problematic (see footnote to Table~\ref{pm_table}).
Given the fact that the TRC proper motion of HR~7169 is consistent with
kinematical membership to the R~CrA association and that HR~7170 presumably
belongs to the same system, it is likely that the different TRC proper motion
of HR~7170 reflects orbital motions within the system.
Although the kinematical membership of HR~7170 could not be proven directly,
it still should be considered an R~CrA association member.
The stars HR 7169 and HR 7170 lie in a cavity in the CO distribution (Loren 1979).
This suggests similar distances to the stars and cloud. 
Additionally, optical images show an extensive reflection nebulosity
surrounding the pair.

The two other stars are the newly identified TTS RXJ1844.3-3541 and RXJ1901.4-3422.
The case for RXJ1844.3-3541 is not clear. According to its spectral type
and lithium line strength, it is clearly a young TTS, and also its radial
velocity is consistent with membership. Either the proper motion from
STARNET maybe in error or the star could have been ejected from the cloud.
Comparing its proper motion with the association mean, it would move
towards the south-west relative to the cloud, but it is located north-west 
of the cloud, so that it probably was not ejected from the R~CrA cloud.

For RXJ1901.4-3422 the classification as TTS is doubtful. At a late-F spectral 
type, even Pleiades ZAMS stars have not yet burned significantly lithium,
so that classification as either pre-MS or ZAMS stars is difficult.
Because its lithium line is stronger than its nearby calcium line,
we classify the star as TTS. The Hipparcos parallax of
65$\pm$5\,pc clearly places it foreground to the R~CrA cloud, which explains
its large proper motion, and it is the only star located clearly north of the
R~CrA cloud. We conclude that RXJ1901.4-3422 is most likely not an
association member, but young.

\section {Completeness of our survey}
\label{comp}

In Fig. 10, we compare the complete log~N--log~S curve 
with those for stellar, extra-galactic, and unidentified sources.
At a PSPC count rate threshold of 0.1~cts~$s^{-1}$, $90\%$ of the
RASS sources have optical counterparts, most of them being stars.
At lower count rates the fraction of unidentified sources
increases from $\sim 30\%$ at 0.07~cts~$s^{-1}$ to more than 
$50\%$ at 0.02~cts~$s^{-1}$, a level at which all curves flatten because of 
the incompletness of the RASS itself. 
Above a PSPC count rate of 0.03 cts s$^{-1}$ we have optically
identified 75 among 136 RASS X-ray sources (i.e. $55\%$), most of them
being stars. The small number of extra-galactic object identified at
this threshold (only two) is simply due to our follow-up observation
strategy.  According to the results of Guillout (1996) and Zickgraf  et
al. (1997), we conclude that the extra-galactic population in the R CrA
region is likely to account for about $25\%$ of the unidentified X-ray
sources detected above 0.03 cts s$^{-1}$, the rest (i.e. $20\%$) 
probably being optically faint active K- and M-type stars.

\begin{figure}
\label{NSpop}
\vbox{\psfig{figure=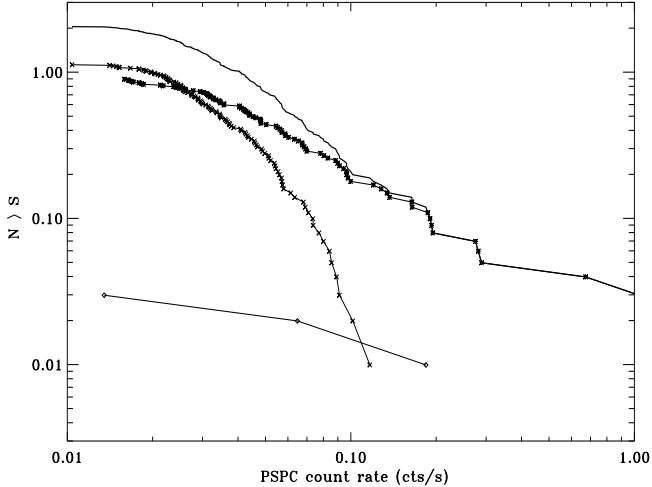,width=9cm,height=7cm}}
\caption{ {\bf The log N--log S curves in the CrA region.}
Number of X-ray sources per deg$^{2}$ detected by ROSAT 
as a function of PSPC count rate for all
RASS sources listed in Table 2 (solid bold line), for the optically
identified stellar (star symbols) and extra-galactic (diamonds)
populations as well as for the unidentified sources (x)}
\end{figure}

We now focus on the identified stellar population and define two
sub-regions within our field, namely the {\it on-cloud} region
(from $\alpha$ = 18h 56m to 19h 24m and from 
$\delta = -38^{\circ}$ to $-36^{\circ}$, i.e. 14~deg$^{2}$) 
and the {\it off-cloud} region
(complementary to the {\it on-cloud} region, i.e. 112~deg$^{2}$).

\begin{figure}
\label{NSste}
\vbox{\psfig{figure=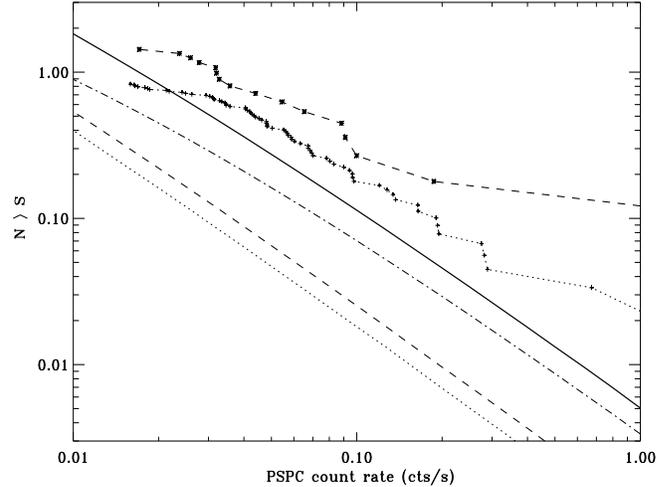,width=9cm,height=7cm}}
\caption{ {\bf The stellar log N--log S curves in the CrA region.}
Number of stellar X-ray sources per deg$^{2}$ optically 
identified {\it on-cloud} (star symbol) 
and {\it off cloud} (plus symbol) regions as well as predictions
from stellar X-ray population model. Bold curve is for all model age
bins, dash-dotted line for stars younger than 150 Myrs, dashed line for
stars from 150 to 1000 Myrs old and dotted line for stars older than
1000 Myrs}
\end{figure}

We have plotted in Fig. 11 the observed {\it on-cloud} and {\it
off-cloud} stellar log N--log S curves as well as the predictions of
the stellar X-ray population model from Guillout et al. (1996).
Computations were run for $|b|$~=~15$^\circ$ and $l$~=~180$^\circ$
although the galactic longitude is irrelevant at the RASS sensitivity.
Results are summarized in Table 7.

\begin{table}
\caption{{\bf X-ray source density.}
Surface density of stellar X-ray sources observed {\it on}
($\rho_{On}$) and {\it off} ($\rho_{Off}$) R CrA cloud, model
predictions  ($\rho_{model}$) as well as observed-prediction ratio as a
function of PSPC count rate threshold $S$ in cts/s.}
\begin{tabular}{clllll} \hline
$S$ & $\rho_{On}$ & $\rho_{off}$ & $\rho_{model}$ &
$\rho_{On}$/$\rho_{model}$ & $\rho_{Off}$/$\rho_{model}$\\
\hline
0.10 &   0.179  &    0.167  &   0.113 & 1.58 & 1.47\\
0.05 &   0.626  &    0.414  &   0.273 & 2.29 & 1.51 \\
0.03 &   1.073  &    0.682  &   0.482 & 2.22 & 1.41 \\ \hline
\end{tabular}
\end{table}

First we note that at any PSPC count rate
the {\it on-cloud} stellar density is significantly higher (by a
factor 2) compared to the model predictions, as expected for 
a region with ongoing star formation. On the other hand,
we expect that the {\it off-cloud} log~N--log~S curve lies within
$15\%$ of the model prediction, which is clearly not the case.

In order to check the relevance of the model prediction, we compare with 
the so-called RasTyc sample (Guillout et al. 1999), a sample of all objects
included in both RASS and Tycho, i.e. the largest sample of stellar X-ray 
sources with homogeneous and accurate data constructed so far. In order 
to account for the magnitude and X-ray flux limited biases of of RasTyc,
we ran a specific model 
optically limited at 10.5~mag (plus $20\%$ of stars down to 11.5~mag). 
We then compared the expected number of stars per deg$^{2}$ with the observed
one computed in two regions extending $10^{\circ}$ wide all around the
sky and centered at $|b|$~=~15$^\circ$. At a PSPC count rate threshold 
of 0.03~cts$s^{-1}$, there are 1819 RasTyc stars detected within these 
two regions amounting to 6946~deg$^{2}$, i.e. 0.26 stars per deg$^{2}$. 
At this level, the model predicts a stellar surface density of 
0.23 stars per deg$^{2}$, in very good agreement with the observations.
We are thus confident that the theoretical log N--log S curves plotted
in Fig. 11 give a good estimation of the ambient galactic plane
stellar population at the R CrA cloud galactic latitude.

We then conclude that in the region surrounding the CrA molecular
cloud our observations reveal $\sim~40\%$ excess of stellar
X-ray sources with respect of a 'pure' galactic plane population
(see Table 7).
According to the expected contribution of extra-galactic sources to
the unidentified population, $40\%$ is a lower limit on the excess.
Such excesses were also detected around other star forming
regions (see Neuh\"auser 1997 and references therein).
However, contrary to some other star forming regions like Lup-Sco-Cen
(Guillout et al. 1998a,b), the Gould Belt can hardly be an explanation 
because of the position of the CrA molecular cloud projected well below the
Gould Belt plane. 

Also around the Chamaeleon clouds (Alc\'ala et al. 1995) and south of the Taurus 
clouds (Neuh\"auser et al. 1997), many new pre-MS stars were found, although
there is no Gould Belt in that directions. As far as the Chamaeleon off-cloud
TTS are concerned, Mizuno et al. (1998) found new, previously unknown, small 
cloud-lets near one third of the off-cloud TTS, which may be the birth places 
of those seemingly off-cloud TTS. If one can explain off-cloud
TTS around the CrA and Cha clouds by cloud-lets rather than by the Gould Belt,
at least some of the Lup-Sco-Cen, and Orion off-cloud TTS may also have
originated in such small cloud-lets, as originally proposed for the Chamaeleon
off-cloud TTS by Feigelson (1996).

The question now is whether we found {\em all} young, i.e.
coronally active stars (inside and) around the CrA dark cloud. 
This can be investigated by optical follow-up observations
of additional unidentified X-ray sources found in deep ROSAT PSPC 
and HRI pointed observations (Walter et al., in preparation).

Whether all young stars were found among {\em all} RASS sources
can be investigated in the following way:
Sterzik et al. (1995) have shown that it is possible to pre-select TTS
candidates from the RASS using four criteria, 
namely the two hardness ratios, the X-ray count rate, and the optical 
magnitude of the nearest (if any) counterpart (within, say, 40").
Then, TTS candidates are those RASS sources which resemble best 
previously known RASS-detected bona-fide TTS according
to the same properties. The parameter which describes how well
a particular RASS source resembles the typical TTS properties
is called {\em discrimination probability} P,
described in detail in Sterzik et al. (1995).

\begin{figure}
\label{mfs}
\vbox{\psfig{figure=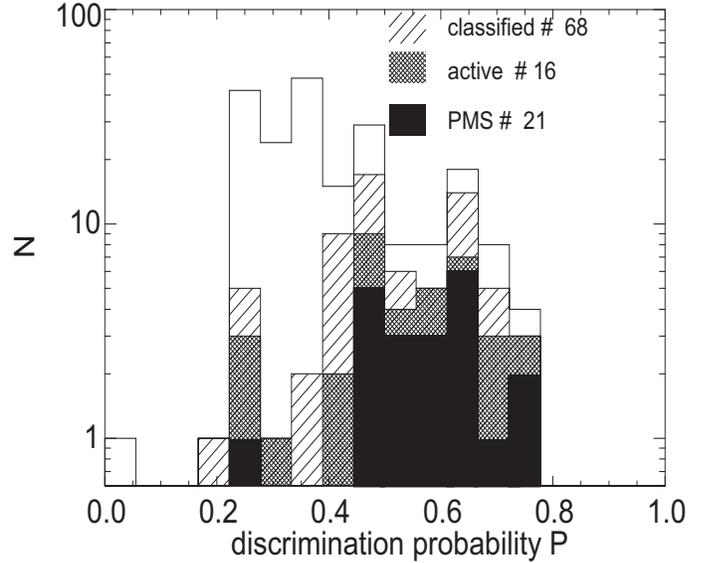,width=9cm,height=7.5cm,angle=90}}
\caption{ {\bf Completeness of our follow-up observations.}
This histogram shows the number of X-ray sources per discrimination 
probability P for PMS stars (17 RASS sources identified with new PMS stars 
plus four RASS sources identified with previously known PMS stars), 
otherwise coronally active stars
(nine ZAMS stars plus seven dKe/dMe stars),
68 X-ray sources identified with other objects,
and all 206 RASS sources in CrA }
\end{figure}

In Fig. 12, we plot the number of CrA RASS sources per 
discrimination probability P, namely for PMS stars, otherwise
active stars, other objects, and unidentified RASS sources.
If we would have pre-selected TTS candidates using the
Sterzik et al. (1995) method, i.e. if we would have done
optical follow-up observations only for RASS sources with a 
discrimination probability of, say, $P \ge 0.45$, we would have 
obtained a high success rate by loosing only one TTS.

Now, for a discrimination probability of, say, $P=0.5$,
the reliability ({\em rel}) of the TTS candidate selection 
is 0.45, based on the classified sub-sample.
The reliability number gives the fraction of real TTS
(real according to our spectroscopy) among those X-ray
sources with discrimination probability above some
threshold, e.g. $P \ge 0.5$.
The fraction of lost unidentified TTS is 0.17,
which is the number of real TTS to be expected (according 
to their discrimination probability values) among those
X-ray sources not observed by optical spectroscopy.
Because there is a total of $N = 46$ sources with $P \ge 0.5$
and 160 below this threshold, the expected number of
TTS hidden in the RASS sample is
\begin{displaymath}
\quad \le~N(P \ge 0.5) \cdot rel + N(P < 0.5) \cdot loss~=~48
\end{displaymath}
Hence, there should be a total of $\le 48$ X-ray sources with TTS
as true counterparts (with an X-ray flux above the RASS flux limit). 
The lower limit can be found when considering only those RASS sources
which are unclassified, but do have an optical counterpart;
there are $\tilde N = 46$ such sources with $P \ge 0.5$ 
and 46 with $P < 0.5$.
Hence, there should be
\begin{displaymath}
\quad \ge~~\tilde N(P \ge 0.5) \cdot rel + \tilde N(P < 0.5) \cdot loss~=~29
\end{displaymath}
RASS sources with TTS as true counterparts.
Of those 29 to 48 sources, 21 are identified as such, namely
as previously known or newly found TTS. The remaining eight to 27
still unknown TTS should be found among as yet unidentified RASS sources.
According to Fig. 10, most of those missing TTS identifications are due 
to the magnitude limit in the catalogs used here (e.g. the GSC).

\section{Summary}

In this data paper, we presented ground-based optical and infrared follow-up observations 
of previously unidentified ROSAT All-Sky Survey X-ray sources in and around the R~CrA
star forming cloud. We identified two new cTTS and 17 new wTTS having stronger lithium
absorption than ZAMS stars of the same spectral type.
Radial velocities and proper motions of most of these new TTS are consistent with
the previously known CrA TTS, but the new objects are distributed more widely in space.
Compared to a galactic model, this sample of ROSAT TTS constitutes an excess
of young stars, more than expected at this galactic latitude. We estimated that 
there should be in total 29 to 48 RASS detectable TTS in CrA, of which 8 to 27 
hitherto unknown TTS are still hidden among as yet unobserved ROSAT Survey sources.
Similar to the two new cTTS found to be located projected onto two small
cloud-lets, the seemingly off-cloud wTTS found here may have formed in
small cloud-lets, which have dispersed since they formed those stars.
For this scenario as well as for the ejection model (run-away TTS), 
a larger than observed velocity dispersion would be expected.
The space motions of the on-cloud and off-cloud TTS are consistent with a
scenario in which cloud and stars oscillate around the galactic plane, with a
very small intrinsic velocity dispersion. This could possibly be the result of
a high-velocity cloud impact triggering star formation, as advocated by
L\'epine \& Duvert (1994). However, alternative scenarios cannot be ruled out
given the accuracy of the kinematical data available at present.

\acknowledgements{
We would like to thank the staff at ESO La Silla and CTIO for their help as well as 
George Leussis for his significant contribution in the reduction of the photometric data.
We are gratefull to our referee, Dr. Bruce Wilking, for many helpful comments,
which improved the content of this paper.
This research has made use of the Simbad database, operated at CDS, Strasbourg, France.
This publication makes use of data products from the Two Micron All Sky Survey, which is 
a joint project of the University of Massachusetts and the Infrared Processing and 
Analysis Center/California Institute of Technology, funded by the National Aeronautics 
and Space Administration and the National Science Foundation.
The ROSAT project is supported by the German Government (BMBF/DLR) and the Max-Planck-Society. 
RN wishes to acknowledge financial support from the Bundesministerium f\"ur Bildung und 
Forschung through the Deutsche Zentrum f\"ur Luft- und Raumfahrt e.V. (DLR) under 
grant number 50 OR 0003.}

{}

\end{document}